\documentclass[twocolumn,prb,aps,showpacs,superscriptaddress,showkeys,reprint]{revtex4-1}

\usepackage{amssymb, amsmath}
\usepackage[english]{babel}
\usepackage{bm}
\usepackage{graphicx}
\usepackage[utf8]{inputenc}
\usepackage{color}

\newcommand\mydots{...\,}

\newcommand{\SANchange}[1]{#1} 

\newcommand{\REVchange}[1]{#1} 

\begin{document}
\author{Sandra C. Kuhn}
\author{Marten Richter}
\email[]{marten.richter@tu-berlin.de}

\affiliation{Institut für Theoretische Physik, Nichtlineare Optik und
Quantenelektronik, Technische Universität Berlin, Hardenbergstr. 36, EW 7-1, 10623
Berlin, Germany}

\title{Tensor network strategies for calculating  biexcitons and trions in monolayer 2D materials beyond the ground state}


\begin{abstract}
Recently in [Phys. Rev. B 99, 241301(R) (2019)] tensor networks build upon logical circuits were briefly introduced
to retrieve exciton and biexciton states. Compared to a conventional approach the tensor network methods scales logarithmic instead of linear in the grid points of the Brioullin zone and linear instead of exponential in the number of electrons and holes. 
This enables calculations with higher precision on the full Brioullin zone than previously possible.
In this paper extensive details for an efficient implementation and the corresponding mathematical background are presented. In particular this includes applications and results for excitons, trions and biexcitons (for monolayer MoS$_2$ as example), going beyond the initial brief introduction.
Furthermore strategies for calculating selective excited bound states and  tests of common approximations are discussed making use of the high accuracy full Brioullin zone treatment of the tensor network method.
 \end{abstract}

\date{\today}
\maketitle
\section{Introduction}
Bound electron hole states such as exciton, trions and biexcitons have been in the focus of semiconductor research since decades \cite{lindberg1988effective,butscher2005ultrafast,doi:10.1002/pssb.200879612,reiter2007spatiotemporal,schmidt2016ultrafast,PhysRevLett.92.067402,winzer2010carrier,PhysRevB.97.035425}. Recently interest was intensified due to the introduction of monolayer two dimensional materials such as transition metal dichalcogenides TMDCs, which show bound electron-hole complexes with very high binding energies \cite{qiu2013optical,berghauser2014analytical,qiu2015nonanalyticity}.
Excitons are routinely calculated in effective mass approximation \cite{haug2009quantum,huser2013dielectric,berghauser2014analytical}, but also on the full Brioullin zone with different accuracy  \cite{ridolfi2018exstruc,deilmann2019finite,steinhoff2018frequency,qiu2013optical,qiu2015nonanalyticity}.

On the other hand calculations of bound electron hole complexes involving a higher number of particles such as trions, biexcitons or charged biexcitons remain sparse \cite{stebe1997excitonic,mayrock1999weak,zimmermanntrion2000,esser2001theory,mostaani2017diffusion,PhysRevB.93.041401,10.1021acs.nanolett.8b00840,doi:10.1021/acs.nanolett.5b03009,PhysRevB.93.125423,Drueppel2017,steinhoff2018biexciton}. Often the theoretical approach for larger complexes rely on approximations: effective mass approximation \cite{stebe1997excitonic,mayrock1999weak,zimmermanntrion2000,esser2001theory,doi:10.1021/acs.nanolett.5b03009}, certain Ansatz functions\cite{mayrock1999weak,katsch2019theory} or restrict the calculation to a part of the Brioullin zone \cite{steinhoff2018biexciton}. A high number of grid points is not feasible for numerical calculations  for  larger complexes.
Main problem are that the many particle wave functions are high dimensional (grid points of the Brioullin zone) and high rank tensors leading to unfeasible memory and computational requirements.

In \cite{Kuhn:2019} it was briefly shown, that  tensor network methods (including matrix product states \cite{vidal2003efficient,orus2014practical,schollwock2011density,CIRAC2017100,verstraete2006matrix,vidal2007classical,CIRAC2017100,plenioheisenberg}, DMRG \cite{white1992density,schollwock2011density} and quantics tensor trains \cite{oseledets2009approximation,oseledets2010approximation,khoromskij2011dlog,kazeev2012low,khoromskaia2015tensor,benner2017fast}) using logical circuits, can be  used to calculate  bound exciton and biexciton states on a Brioullin zone with a high number of grid points.

However the first introduction of the method in Ref. \onlinecite{Kuhn:2019} was targeted to general cluster/correlation expansion treatments and no discussion of the details and pitfalls for creating an efficient implementation were given. Furthermore  strategies to selectively calculate a specific excited bound eigenstates are a missing ingredient, since most tensor network methods such as DMRG often retrieve only the ground state of the system and often the ground state is a rather uninteresting dark state.

Therefore this paper gives  a very detailed introduction into the method specialized to wave functions of electron hole complexes. This includes also details of the best gauge for the encoding of band structure, tight binding coefficients and Coulomb coupling into the tensor network with logical circuits and also ideas for the transformation into matrix product operator (MPO) form are discussed.
After a recapitulation of DMRG and imaginary time propagation as methods for retrieving eigenstates of the Hamilton operator, exciton, trion and biexciton eigenstates are calculated and discussed for a MoS$_2$ monolayer material.
Focus is on retrieving selected excited eigenstates beyond the ground state, accordingly coordinate transformations and projections into parts of the Brioullin zone applied directly on the tensor level are discussed as well as approaches for a faster calculation using imaginary time propagation. 
\SANchange{ Readers only interested in physical results can  skip to Sec. \ref{exciton_section} and following.}

\section{Model system, Hamiltonian and generalized Wannier equation}
\begin{figure}[tb] 	
  \centering
    \includegraphics[width=8.0cm]{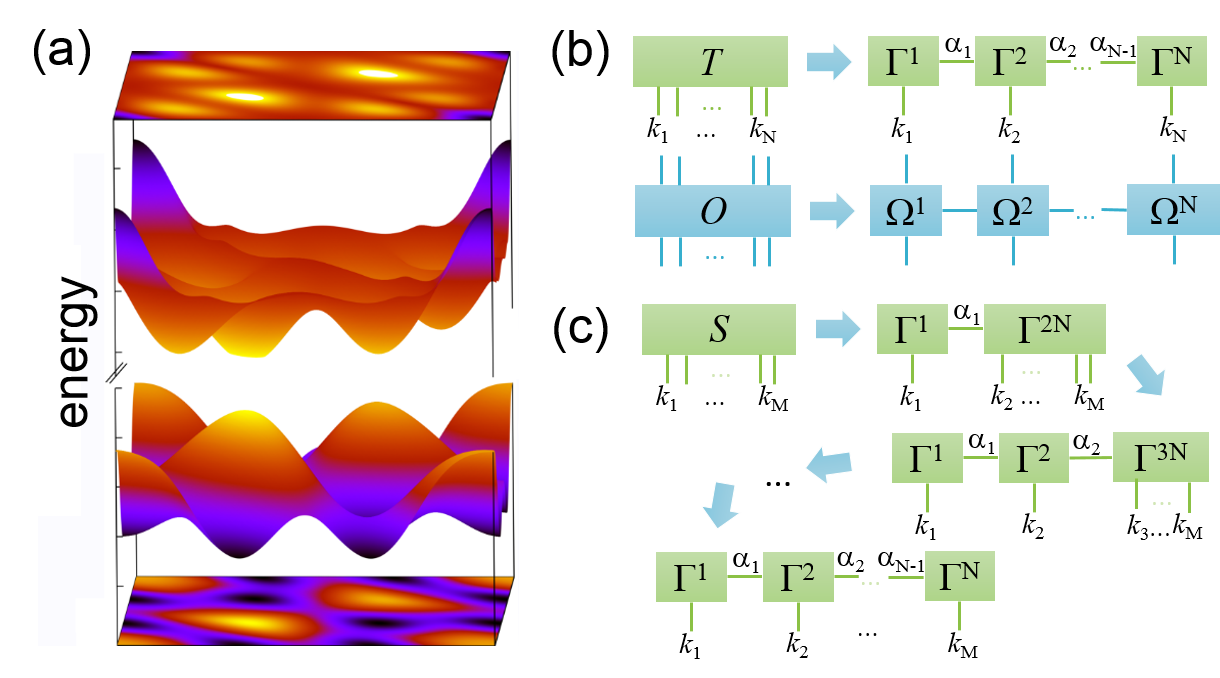}
    \caption{ (a) Valence and conduction band of MoS$_2$ (single band) plotted over the Brioullin zone, (b) decomposition into MPS and MPO and (c) illustration of typical tensor decomposition using SVD. }
    \label{bandtendecomp}
  \end{figure}

The paper is focused on method development, a realistic but slightly simplified model is used with a similar complexity as the full system.
The Hamiltonian $H=H_0+H_C$ for bound electron hole complexes in semiconductors includes the band structure $H_0$ and the Coulomb interaction $H_C$.
The electronic bandstructure $H_0$ reads:
\begin{eqnarray}
H_0=\hbar\sum_{\mathbf{k}\lambda}\varepsilon_{\mathbf{k}}^\lambda a^\dagger_{\mathbf{k}\lambda} a_{\mathbf{k}\lambda},
\end{eqnarray}
where $a^\dagger_{\mathbf{k}\lambda}$ and  $a_{\mathbf{k}\lambda}$  are the creation and annihilation operators for an electron (conduction bands) or hole (valence bands) with the band index $\lambda=c,v$  and the quasi momentum $\mathbf{k}$). 
 $\lambda$  includes as a multiindex also the spin $\sigma$.
 For a valence band $\lambda=v$ it holds $a^\dagger_{\mathbf{k}\lambda}=e^\dagger_{\mathbf\lambda}$  as well as for the conduction band $\lambda=c$ $a^\dagger_{\mathbf{k}\lambda}=h^\dagger_{\mathbf\lambda}$.  Here $\varepsilon_{\mathbf{k}}^\lambda$ represents the band structure for the respective carrier, where we restrict our discussion to the valence and conduction band (including spin, cf. Fig. \ref{bandtendecomp})  for  the electron hole complexes. For the numerical results presented here, we use the tight binding bandstructure from \cite{ridolfi2015tight,ridolfi2018exstruc}. 
For the numerical calculation the quasi momentum $\mathbf{k}$ is  $\mathbf{k}=1/\tilde{N}(k_x \mathbf{b}_x + k_y \mathbf{b}_y)$  using the basis vectors $\mathbf{b}_x$ and $\mathbf{b}_y$ of the two dimensional Brioullin zone ($\mathbf{b}_{x/y}$ are not the Cartesian basis elements) with $k_{x/y}$ being integer numbers and $\tilde{N} $ the number of grid points in one dimension resulting in a Monkhorst-Pack grid.
For the  presented method, the quasi momentum is written using the bits $k_{x/y}^{(i)}$ of the integer numbers 
\begin{eqnarray}
\mathbf{k}=\frac{1}{2^N} \left(\sum_{i=1}^N (k_x^{(i)} \mathbf{b}_x + k_y^{(i)} \mathbf{b}_y)2^i-k_x^{shift} \mathbf{b}_x -k_y^{shift} \mathbf{b}_y\right). \label{bitrepresentation}
\end{eqnarray}
$N$ is the number of bits per dimension and $k_{x/y}^{shift}$ are integer numbers, that are used to shift the Brioullin zone grid, and will be beneficial for applications discussed later using the tensor representation.
Furthermore note, that many properties of the semiconductor are periodic with respect to the reciprocal lattice vector $\mathbf{G}$, so that the index $\mathbf{k}$ and $\mathbf{k}+\mathbf{G}$  label often the same quantity (valid e.g. for band structure tight-binding coefficients etc., not valid for Coulomb potential).
The Coulomb interaction Hamiltonian reads:
\begin{eqnarray}
H_C=\sum_{\mathbf{k}_1\mathbf{k}_2\mathbf{q}\lambda_1\lambda_2 } I^{\mathbf{k}_1\mathbf{k}_2\mathbf{q}}_{\lambda_1\lambda_2 } a^\dagger_{\mathbf{k}_1\lambda_1} a^\dagger_{\mathbf{k}_2\lambda_2}
a_{\mathbf{k}_2+\mathbf{q}\lambda_2}
a_{\mathbf{k}_1-C^{\lambda_1}_{\lambda_2}\mathbf{q}\lambda_1}
\end{eqnarray} 
$C^{\lambda_2}_{\lambda_1}$ is 1, if $\lambda_1$ and $\lambda_2$ are both electrons or both holes and $-1$ otherwise.  The coupling element $I^{\mathbf{k}_1\mathbf{k}_2\mathbf{q}}_{\lambda_1\lambda_2 }=F^{\mathbf{k}_1\mathbf{k}_2}_{\mathbf{q}} V_{\mathbf{q}}C^{\lambda_1}_{\lambda_2}$ includes the Coulomb potential $V_{\mathbf{q}}$ (here for demonstration the simple Keldysh style Coulomb potential \cite{PhysRevB.88.045318,berghauser2014analytical}) and the tight-binding (TB) coefficients $c_{\mathbf{k}n}$ inside $F^{\mathbf{k}_1\mathbf{k}_2}_{\mathbf{q}}=\sum_{n_1,n_2} c^*_{\mathbf{k}_1 n_1} c^*_{\mathbf{k}_2 n_2} c_{\mathbf{k}_2+\mathbf{q} n_2} c_{\mathbf{k}_1-C^{\lambda_1}_{\lambda_2}\mathbf{q} n_1}$ including only the long range part of the Coulomb interaction and neglecting short range contributions such as exchange splitting \cite{PhysRevB.47.4569} for the simplified model system.
In the following,  $\mathbf{k}$ is a multiindex $\mathbf{k}={\mathbf{k}\lambda}$ including also the band index $\lambda$ and $\lambda$ is only written separately, if necessary.
The tight binding coefficients originate from the Ansatz of the electronic wave function \cite{ridolfi2015tight}:
\begin{eqnarray}
\psi_{\mathbf{k}\lambda}(\mathbf{r})=\sum_{\mathbf{R}}e^{i\mathbf{k}\cdot\mathbf{R}} \sum_{n_1} c_{\mathbf{k}n_1} \varphi_{n_1,\mathbf{R}}(\mathbf{r}), \label{tbansatz}
\end{eqnarray}
with the lattice vector $\mathbf{R}$ and the tight binding orbitals $\varphi_{n_1,\mathbf{R}}(\mathbf{r})$. The tight binding model from \cite{ridolfi2015tight,ridolfi2018exstruc} is used as basis for the calculation, however the form of Eq. (\ref{tbansatz}) is furthermore compatible with Wannier interpolation \cite{RevModPhys.84.1419}, so that in the future this method may be applied to output from Wannier90 \cite{MOSTOFI2008685} (planned future work). 
 
In this paper methods for calculating the bound eigenstates of electron-hole complexes are presented. Basis for the calculations is a generalization of the Wannier or Bethe Salpeter equation:
\begin{eqnarray}
  && E \,\Psi_{\mathbf{k}_1\mydots \mathbf{k}_n}= \sum_{j=1}^n \varepsilon_{\mathbf{k}_j}
 \Psi_{\mathbf{k}_1\mydots\mathbf{k}_j \mydots \mathbf{k}_n}\nonumber\\
  &&\quad +2\sum_{i<j\mathbf{q}}
  I_{\lambda_j\lambda_i}^{\mathbf{k}_j\mathbf{k}_i\mathbf{q}}/\hbar
  \Psi_{\mathbf{k}_1\mydots \{\mathbf{k}_{i}+\mathbf{q}\}\mydots \{\mathbf{k}_{j}-C^\lambda_{\lambda_j}\mathbf{q}\}\mydots\mathbf{k}_n}, \label{bethesalpeterlike}
\end{eqnarray}
an eigenvalue problem for energy $E$ and the wavefunctions in reciprocal space $\Psi_{\mathbf{k}_1\mydots \mathbf{k}_n}$. 
The wavefunction $\Psi_{\mathbf{k}_1\mydots \mathbf{k}_n}$ are the expansion coefficients for the many particle state $|\Psi\rangle=\sum_{\mathbf{k}_1\mydots \mathbf{k}_n } \Psi_{\mathbf{k}_1\mydots \mathbf{k}_n} a^\dagger_{k_1}\mydots a^\dagger_{k_n}|\psi_0\rangle$. Here, $|\psi_0\rangle$ is the ground state of the system with all valence bands occupied (no holes) and all conduction bands unoccupied (no electrons), so that $(H_0+H_C)| \psi_0\rangle=0$ holds.
It is possible with this ansatz to calculate  charged electron-hole complexes like trions or charged biexcitons \cite{hao2017neutral} for low doping. However higher doping  requires a modified ground state $|\psi_0'\rangle$ and a transformation of the Hamiltonian, so that $(H_0'+H_C')|\psi_0'\rangle=0$.

\section{Matrix product states and product operators}
The wave function $\Psi_{\mathbf{k}_1\mydots \mathbf{k}_n}$ and also the operators acting on the wave function in Eq. (\ref{bethesalpeterlike}) are stored in memory during the solution of the eigenproblem, mapping the quasi momenta $\mathbf{k}_i$  onto a Monkhorst-Pack grid\cite{PhysRevB.13.5188} of the Brioullin zone.
For high accuracy with a $1024\times1024$ grid, the wave functions $\Psi_{\mathbf{k}_1}$ for a single particle as well as bright excitons $\Psi_{e \mathbf{k}_1 h \mathbf{k}_1}$  are  tensors with $\approx 10^6$ elements, resulting in a feasible eigenproblem.
But electron hole complexes with higher particle numbers  result in more indices such as all excitons $\Psi_{h \mathbf{k}_1 e \mathbf{k}_2}$ ($\approx 10^{12}$ elements), the trion  $\Psi_{h \mathbf{k}_1 e \mathbf{k}_2 e \mathbf{k}_3}$ ($\approx 10^{18}$ elements) or the biexciton $\Psi_{h \mathbf{k}_1 e \mathbf{k}_2 h \mathbf{k}_3 e \mathbf{k}_4}$ ($\approx 10^{24}$ elements), so that
the eigenproblem is not feasible anymore because of the high memory requirement. In general the overall number of tensor elements for $\Psi_{\mathbf{k}_1\mydots \mathbf{k}_n}$   scales as $g^n$ with $g$ the number of grid points and $n$ the number of particles.
So that the calculation of higher electron/hole complexes such as trions and biexcitons
was always carried out on grids with fewer points or a restriction to parts of the Brioullin zone \cite{Drueppel2017,steinhoff2018biexciton}.
To overcome this issue we proposed in Ref. \onlinecite{Kuhn:2019} to use matrix product states (MPS) \cite{vidal2003efficient,orus2014practical,schollwock2011density,CIRAC2017100} (also  called tensor trains in mathematics \cite{oseledets2009approximation,oseledets2010approximation,khoromskij2011dlog,kazeev2012low,khoromskaia2015tensor,benner2017fast}) to represent the clusters.
MPS were introduced in \cite{vidal2003efficient}, where it was shown, that every tensor $T_{k_1,\dots,k_N}$ can be approximated by a MPS as
\begin{eqnarray}
T_{k_1,\dots,k_n}=\sum_{\alpha_1,..,\alpha_{n-1}} \Gamma^{1,k_1}_{\alpha_1}\Gamma^{2,k_2}_{\alpha_1\alpha_2}\Gamma^{3,k_3}_{\alpha_2\alpha_3}
   \cdots \Gamma^{n,k_{n}}_{\alpha_{n-1}}.
\end{eqnarray}   
 Here, the tensors $\Gamma^{n,k}_{\alpha\alpha'}$ have less than $g\cdot {D}^2$ elements with  $D$  the maximum dimension of  $\alpha_i$ (link dimension).
  If the maximum link dimension $D$ is sufficiently small for an accurate representation of the tensor $T$, 
  the required memory does not scale exponentially anymore $g^{n}$ in the number of particles $n$   but linear $n\cdot g\cdot{D}^2$ enabling 
  a  calculation using the high rank tensors.
  In the following, we will use the common graphical notation for tensors, where each tensor is represented by a box and each index is depicted by a line with 
  connected lines representing contracted indices (cf. Fig. \ref{bandtendecomp}).
  Tensors are mathematically vectors and typical vector operations such as addition, subtraction, norm, scalar multiplication are directly calculated in MPS form.
  Also linear operations acting on tensors can be represented as matrix product operators (MPO) (see Fig. \ref{bandtendecomp}b)), for which efficient applications directly to MPS exists \cite{schollwock2011density}.
  Instead of using  the indices for the quasimomentum $\mathbf{k}$, we use the binary representation $k_{x/y}^{(i)}$ of the Monkhorst-Pack grid from Eq. (\ref{bitrepresentation}) as the indices used for the representation in MPS form. This further reduces the scaling to $n\cdot \log(g)\cdot D^2$, so that the number of grid points $g$ only effect logarithmically the number of tensor elements. This idea comes originally from quantics tensor trains introduced in \cite{oseledets2009approximation,oseledets2010approximation,khoromskij2011dlog,kazeev2012low,khoromskaia2015tensor,benner2017fast}, here it is also beneficial, since it enables the calculation with logical circuits \cite{Kuhn:2019}.
\REVchange{ Also the bit decompositions results in a hierarchical decomposition of the BZ, since for every dimension the first bit divides the BZ into an upper and lower part, and the next bits subsequently divide these parts into two halves. }  

Important for  efficient MPOs constructed using the logical circuits is the ordering of the bits in the MPS.
Bits, which are connected via logical circuits, should not be far away in the MPS in order to achieve a low link dimension in the MPO.
For the tensor $\Psi_{\mathbf{k}_1\mydots \mathbf{k}_n}$ we will use the bit ordering ($[eh]$, $[\lambda]$, $[k^{(1)}_x]$, \dots, $[k^{(N)}_x]$, $[k^{(1)}_y]$, \dots, $[k^{(N)}_y]$) for the band index, spin index, and the $\mathbf{k}$ indices. Here $[\,\cdot\,]$ is a group of bits: $[\Lambda]=  \Lambda_1,\dots,\Lambda_n$, cf. Fig. \ref{tnthomogen} a). 

\REVchange{
Note, that the setup used here is very different to two dimensional systems typically treated with tensor networks or MPS such as two dimensional spin lattice etc.. Our system is treated in reciprocal space. It is a periodic system with only few atoms in the elementary cell restricted to only the optical relevant valence and conduction bands.  While traditionally the indices used for MPS decompositions are connected to a site (a quantum system located at a specific spatial position), here the bit representation of quantum numbers of wave functions such as the quasi moment is used for the decomposition.
Therefore it is unclear, if this setup suffers from the same exponential scaling known from two dimensional spin lattice \cite{RevModPhys.82.277} and thus special care and  a thorough convergence analysis  is required and was done (see supplemental material of Ref.\onlinecite{Kuhn:2019} and Sec. \ref{convergentexcited} for details).
}

 \section{Converting matrix properties to MPS}
 \begin{figure}[tb] 	
 \centering
   \includegraphics[width=6.0cm]{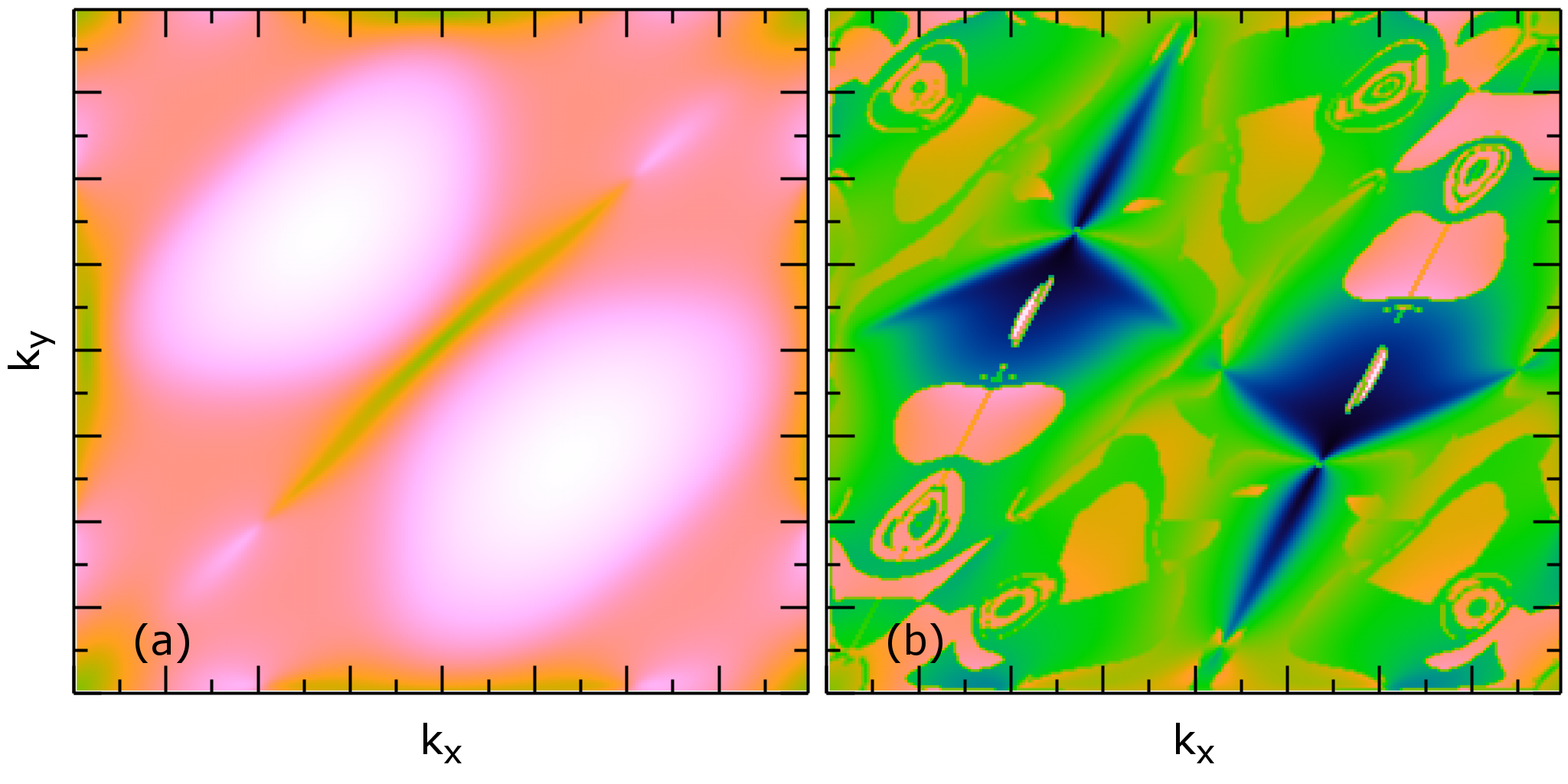}
   \caption{ Real part of a tight binding coefficient $c_{\mathbf{k}\lambda,n}$  (a) with phase correction and (b) without phase correction plotted over the Monkhorst-Pack grid. (a) is smooth and shows symmetries and (b) shows pronounced discontinuities and pattern, which increase the required link dimension.   }
   \label{tbcoefffig}
 \end{figure}
A necessary step for converting  Eq. (\ref{bethesalpeterlike})  into MPO form is the conversion of the coefficients from Eq. (\ref{bethesalpeterlike}).
We start with a discussion of the bandstructure $\varepsilon_{\mathbf{k}}$ and the closely related tight binding coefficients $c_{\mathbf{k},n}$ (Note: similar quantities such as dipole moments are transformed in an analogous way).
The first step is to store the whole tensor, e.g. $\varepsilon_{\mathbf{k}}$ in memory and then to recast it as $\varepsilon_{\lambda,k^{(1)}_x,\mydots,k^{(n)}_x,k^{(1)}_y,\mydots,k^{(n)}_y}$ with the bits as indices.
The tensor $\varepsilon_{\lambda,k^{(1)}_x,\mydots,k^{(n)}_x,k^{(1)}_y,\mydots,k^{(n)}_y}$ is then converted using successive application of singular value decompositions (SVD) following the usual algorithm \cite{vidal2003efficient} as illustrated in Fig. \ref{bandtendecomp} c).
However the resulting MPS has a high link dimension, so that a compression using a variational algorithm \cite{schollwock2011density} is carried out in the next step to reduce the link and to speed up later calculations. 
The same procedure is also used for the tight-binding coefficients $c_{\mathbf{k}\lambda,n}$, where only tight binding orbit index $n$ is not decomposed into bits $c_{\lambda,k^{(1)}_x,\mydots,k^{(n)}_x,k^{(1)}_y,\mydots,k^{(n)}_y,n}$.

The tight binding coefficients $c_{\mathbf{k}\lambda,n}$ contain much more information than the bandstructure $\varepsilon_{\mathbf{k},\lambda}$  and require  higher link dimensions for accurate results.  
Therefore we  developed strategies to primarily reduce the link dimension of the compressed MPS for the tight binding coefficients.
In our first attempts we tried to shift the origin of the Brioullin zone using $k^{shift}_x$ and $k^{shift}_y$.
The idea was  that depending on the origin symmetries in the MPS bit representation are different and  may lead to a lower link dimension by maintaining the same accuracy.
While we saw some changes in the link dimension (may be a factor 2 or 3) before the compression, but not consistent for all quantities, the compression using the variational algorithm after the first decomposition has shown to be more effective.
Instead  a shift of the Brioullin zone may be used to restrict the calculation to certain parts of the Brioullin-Zone using bit masks (see discussion in Section \ref{exciton_section}).

For achieving  a low link dimension, the data has to be smooth and symmetric with respect to $\mathbf{k}$.
For example, in our first implementations the wrong bands were selected in the vicinity of points in the Brioullin zone with band crossings, while these points were energetically far away from the important K and K' points and should not alter the results, the increased link dimensions were prohibitive.
Therefore special care was taken when assigning an eigenvectors  eigenvalue from the tight-binding eigenvalue problem \cite{RevModPhys.84.1419} of the form:
\begin{eqnarray}
\sum_m H_{\mathbf{k},nm} c_{\mathbf{k},m} = \varepsilon_{\mathbf{k}} c_{\mathbf{k},n} \label{TB_eigenvalue_problem}
\end{eqnarray} 
to a specific band's energy $\varepsilon_{\mathbf{k}}$ and tight binding coefficient $c_{\mathbf{k},n}$.
One example of a successful strategy is to select an eigenvalue at a certain $\mathbf{k}$-point, whose tight binding coefficients changed least compared to the $\mathbf{k}$-points in the vicinity.
Another problem is related to the phase of the tight binding coefficients, which is as usual completely arbitrary.
Since the eigenvalue problem Eq. (\ref{TB_eigenvalue_problem}) is smooth in $\mathbf{k}$, the eigenvalue solvers show for most parts also a smooth behavior of the phase, however at certain boundaries abrupt changes of the phase  occur (cf. Fig. \ref{tbcoefffig} (a)). 
The jumps lead again to a high link dimension and need to be removed from the data before conversion to MPS.
Therefore   a gauge was applied that fixes the phase of one dominant orbital for a certain band and thus removes random phase jumps (cf. Fig. \ref{tbcoefffig}). 

One might be tempted to include only the tight binding coefficients or bandstructure around the important symmetry points or in a certain energy window to reduce the information and thus the MPS link dimension. However all our attempts leaded  to a significant increase in the link dimension, since the shape of the important area is encoded as well and dominates the required link dimension.

For the Coulomb potential $V_{\mathbf{q}}$, we use the Keldysh Coulomb potential
$V_{\mathbf{q}}={e^2}/(2\varepsilon_0\varepsilon_d q(1+r_0 q))
$
for our model system. Here $\varepsilon_d=(\varepsilon_1+\varepsilon_2)/2$ is the mean value between the surface and air.
The Coulomb potential $V_{\mathbf{q}}$ does not depend on a quasi momentum inside the Brioullin zone, but on a difference between quasi momentums.
If we would just convert $V_{\mathbf{q}}$ to a MPS using the  bit representation of the Brioullin zone, $V_{\mathbf{q}}$ for negative $q_x$ and/or $q_y$ would not be included.
Instead $\tilde{V}_{\mathbf{q}}=V_{\mathbf{q}}+V_{\mathbf{q}+\mathbf{b}_x}+V_{\mathbf{q}+\mathbf{b}_y}+V_{\mathbf{q}+\mathbf{b}_x+\mathbf{b}_y}$ is used as basis for the compression. This exploits the periodicity of the Brioullin-Zone and the representation of negative binary numbers in two's complement notation \cite{tietze2015electronic}.
The further conversion to MPS is than done analog to the bandstructure etc. using the bit representation $\tilde{V}_{q^{(1)}_x,\mydots,q^{(n)}_x,q^{(1)}_y,\mydots,q^{(n)}_y}$. 

\section{MPO constructed using logical circuits}
For calculating the eigenstates $\Psi_{\mathbf{k}_1\mydots \mathbf{k}_n}$  of Eq. (\ref{bethesalpeterlike}), the equation is translated into tensor-network form.
Eq. (\ref{bethesalpeterlike}) is a standard eigenvalue problem:
\begin{eqnarray}
E \Psi_n=\sum_{m} H_{nm} \Psi_m,
\end{eqnarray}
with the vector components of the wavefunction $\Psi_m$ using a generalized index $m$ and a Hilbert operator matrix element $H_{nm}$. Naively viewed every summand of the rhs  takes the component $\Psi_m$ with the index $m$ multiplies it with the matrix element $H_{nm}$ with the indices $n$ and $m$ and adds it to the resulting vector component with index $n$.
So for the tensor network, we must ensure, that from the MPS for $\Psi_{\lambda,k^{(1)}_x,\mydots,k^{(n)}_x,k^{(1)}_y,\mydots,k^{(n)}_y,n}$ on the rhs the correct indices (bits) are connected to the indices of final MPS (lhs), after applying the tensor network. Furthermore the matrix element connected to the initial and final indices has to be also multiplied to the final value by the tensor network.
The procedure is first illustrated for the band energy term in Eq. (\ref{TB_eigenvalue_problem}): $ \varepsilon_{\mathbf{k}_j}
 \Psi_{\mathbf{k}_1\mydots\mathbf{k}_j \mydots\mathbf{k}_n}$, the initial and final index are the same. Also  the index for particle $j$ is connected to $\varepsilon_{\mathbf{k}_j}$. 
 Therefore we can rewrite the term as $\sum_{\mathbf{k'}_j \mathbf{k''}_j}  \delta_ {\mathbf{k}_j \mathbf{k'}_j \mathbf{k''}_j }\varepsilon_{\mathbf{k'}_j}
  \Psi_{\mathbf{k}_1\mydots \mathbf{k}_j\mydots \mathbf{k}_n}$.
Thus $\delta_{\mathbf{k}_j \mathbf{k'}_j \mathbf{k''}_j }$ translates into a product of bitwise $\delta$'s $\Pi_{i} \delta_{\mathbf{k}_j^{(i)} \mathbf{k''}_j^{(i)}  \mathbf{k''}_j^{(i)}  }$ for every k bit including band and spin.
\begin{figure}[tb] 	
  \includegraphics[width=8.0cm]{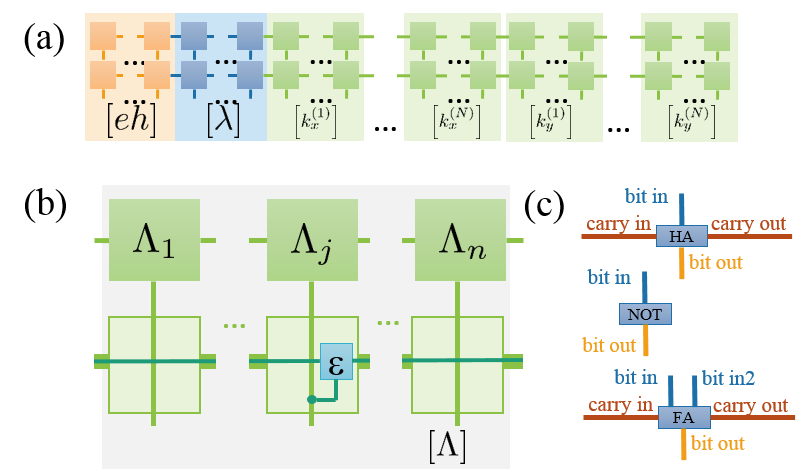}
  \caption{a) General form of MPS and MPO including the ordering of the different bit groups [eh] band index, [$\lambda$] spin, and [$k^{(i)}_{x/y)}$] bits, b) tensor network for band energy part including the MPS $\varepsilon$ for the band structure, c) logical circuits and their inputs. }
  \label{tnthomogen}
\end{figure} 

\begin{figure}[tb] 	
\centering
  \includegraphics[width=8.5cm]{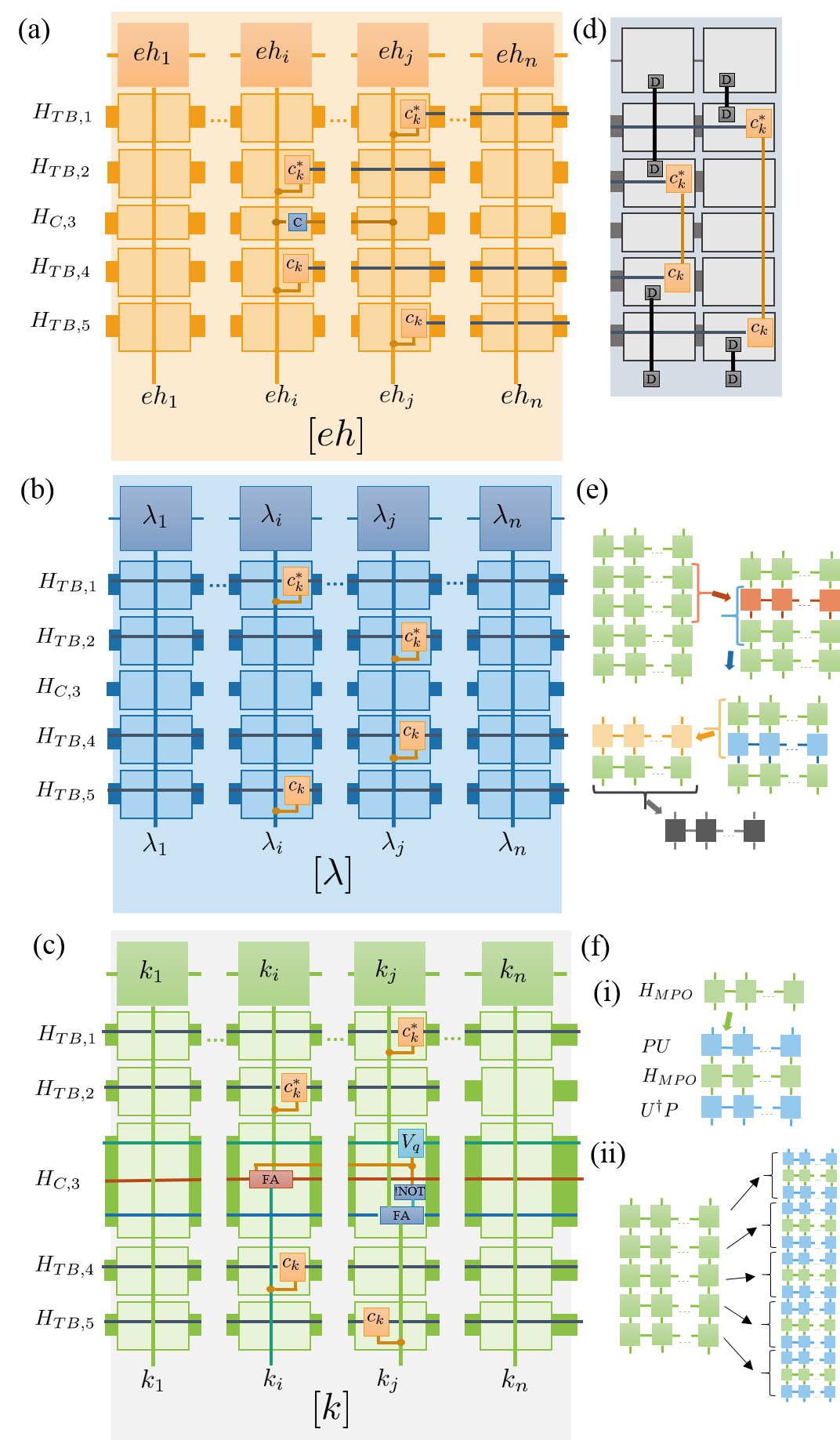}
  \caption{Tensor network representing the Coulomb interaction in Eq. (\ref{bethesalpeterlike}), with a) the band part [eh] b) band index, spin part [$\lambda$] and the c) [$k^{(i)}_{x/y)}$] bit part and (d) the additional tensors for the tight binding summation indices $n_1$ and $n_2$. The tensor $D$ used for termination are normalized to 1, e) efficient multiplication sequence of the Coulomb term MPO representation. f) Application of projector $P$ and unitary transformation $U$ for confining  and transforming coordinates (i) for a single MPO and (ii) for a stacked MPO. }
  \label{tntcoulomb}
\end{figure} 

The corresponding tensor network  is depicted in  Fig. \ref{tnthomogen} b), for $j$ bit $\delta_{\mathbf{k}_j^{(i)} \mathbf{k'}_j^{(i)}  \mathbf{k''}_j^{(i)}  }$ is written as a round dot, which connects the incoming $\mathbf{k''}_j^{(i)}$ index (connected to the lhs MPS on the top), outgoing index $\mathbf{k}_j^{(i)}$ (resulting in the rhs) and index $\mathbf{k'}_j^{(i)}$ connected to tensor of corresponding bit index of the band structure  $\varepsilon_{\mathbf{k}_j}$ MPS. 
For all other bits for particles $l\neq j$ just the ingoing and outgoing bit indices remain the same ($\delta_{\mathbf{k}_l^{(i)} \mathbf{k''}_l^{(i)}  }$), which is written as a straight line.
The tensor network is rather simple and can be represented by a single MPS, since it is diagonal in all bit indices.  
Also we see already the similarity of the tensor network to logical circuits.  

The Coulomb interaction contribution $ +2\sum_{i<j\mathbf{q}}
  I_{\lambda_j\lambda_i}^{\mathbf{k}_j\mathbf{k}_i\mathbf{q}}
  \Psi_{\mathbf{k}_1\mydots \{\mathbf{k}_{i}+\mathbf{q}\}\mydots \{\mathbf{k}_{j}-C^\lambda_{\lambda_j}\mathbf{q}\}\mydots\mathbf{k}_n}, $  in Eq. (\ref{bethesalpeterlike}) is non-diagonal in the indices.
 It connects the incoming indices on the rhs  for particle $i$ $\{\mathbf{k}_{i}+\mathbf{q}\}$  and particle $j$   $\{\mathbf{k}_{j}-C^{\lambda_i}_{\lambda_j}\mathbf{q}\}$ (top of Figure \ref{tntcoulomb}c))   to the outgoing indices $\mathbf{k}_{i}$ and $\mathbf{k}_{j}$ (bottom of Figure \ref{tntcoulomb}c)).
 This is implemented for every bit of the indices  by two tensors, which  are full adder (FA) logical circuits \cite{tietze2015electronic,Kuhn:2019}. The carry output  bit of every full adder for particle $i$ ($j$ analog) is connected to the carry bit input of the full adder handling the next bit (cf. Fig.\ref{tnthomogen}c) ). 
 The input bit of the FA handling the $\mathbf{q}$ related bits are connected to the MPS representation of $V_{\mathbf{q}}$ (cf. Fig.\ref{tnthomogen}c) ).
 In case of a subtraction of $\mathbf{q}$ a NOT circuit is put on the $\mathbf{q}$ bit input of the full adder (cf. Fig.\ref{tnthomogen}c) ).
 Note, that $V_{\mathbf{q}}=V_{-\mathbf{q}}$ was exploited in the implementation.
 The first carry bit input of every dimension is set to $0$ for an addition and $1$ for a subtraction of $\mathbf{q}$, the last carry bit output is set to $0$ and $1$ by a tensor ensuring the periodic boundary conditions of the Brioullin Zone.
The index handling and the multiplication of $V_{\mathbf{q}}$ is handled in the middle MPO $H_{C,3}$ in Fig. \ref{tntcoulomb}c).
The remaining parts of $ I_{\lambda_j\lambda_i}^{\mathbf{k}_j\mathbf{k}_i\mathbf{q}}$ are $c^*_{\mathbf{k}_{i}+\mathbf{q} n_1}$ ($H_{TB,1}$) and $c^*_{\mathbf{k}_{j}-C^{\lambda_i}_{\lambda_j}\mathbf{q} n_2}$  ($H_{TB,2}$) for the incoming indices and $c_{\mathbf{k}_{i}+ n_1}$ ($H_{TB,4}$) and $c_{\mathbf{k}_{j} n_2}$ ($H_{TB,5}$) for the outgoing indices. 
In this case the tight binding coefficient are diagonal in the bitindices and are each handled by one MPO, two on the top for the incoming indices (cf. Fig. \ref{tntcoulomb} c)) and two on the bottom for the outgoing indices. 
Therefore index transformations can be applied to each of the MPOs individually, which we will use later in Sections \ref{exciton_section},\ref{trion_section},\ref{biexciton_section}.
The indices $n_1$ and $n_2$ are added to the end  of the initial MPS and removed for the final MPS after the MPOs were applied (cf. Fig. \ref{tntcoulomb}d)) by merging with adjacent tensors inside the MPS for handling the sum over $n_1$ and $n_2$.
One might wonder, why we organize the tensor network into five stacked MPOs instead of a single MPO. The answer is simple, let us assume that the link dimension of the tight binding coefficients and Coulomb potentials is roughly $D$ a representation using one MPO will scale as $D^5$, which is not feasible, since $D$ is typical in the order of 100.
Therefore a strategy for merging the MPO will be discussed in the next section.

In the actual implementation, we generate two sets of  MPOs one for the case $\lambda_i=\lambda_j$ and one for the case $\lambda_i\neq\lambda_j$. This reduces the link dimension by a factor of 2 and thus the polynomial computing time.

At the end with including the band energy and Coulomb interaction a set of MPOs was generated that represents all necessary terms. In theory, the MPOs could be added before the eigenvalues are calculated. However the summed MPOs have a much higher link dimension D than the individual, so that it is for most algorithms more efficient to not sum the MPOs, but the resulting MPS (imaginary time propagation) or the environment tensors \cite{itensor}.

\subsection{Merging MPOs into a single MPO}
\begin{figure}[tb] 	
\centering
  \includegraphics[width=7.0cm]{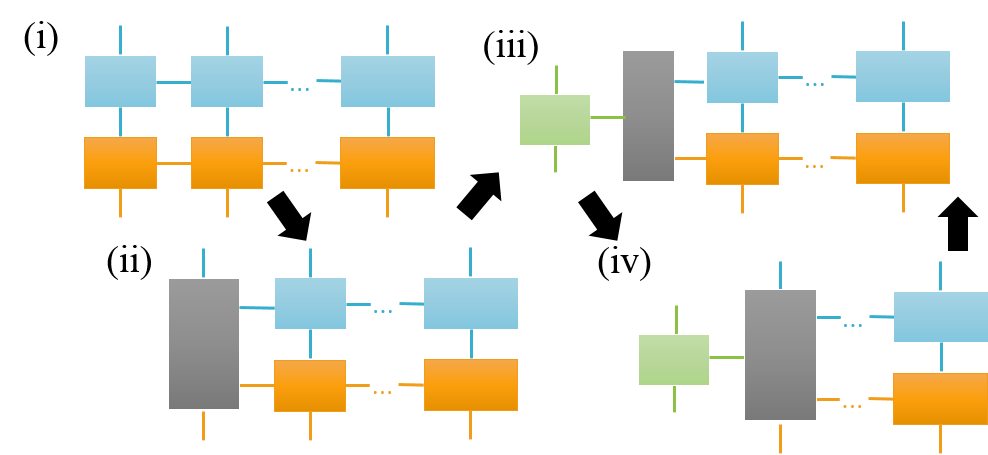}
  \caption{ Typical algorithm for merging two MPOs: (i) Starting by multiplying the first two tensors, the resulting tensor (ii) is decomposed by a SVD (iii) then (iv) the steps are repeated by merging the next tensors of the MPO. The end the resulting MPO is orthogonalized. }
  \label{nmultfig}
\end{figure} 
The five stacked MPOs of the Coulomb interaction have to be merged into a single MPO for algorithms like DMRG.
The typical algorithm for multiplying two MPOs \cite{schollwock2011density,itensor} (cf. Fig. \ref{nmultfig}) goes from left to right through the orthogonalized MPOs multiplies the tensor of the two MPOs and also does a SVD at every tensor. Afterwards the resulting MPO is orthogonalized.

Naively, the stacked MPOs can be multiplied from top to bottom to get the resulting merged MPO, but this is very inefficient and for realistic intermediate link dimensions (limited to a feasible range of a few thousands) the resulting MPO will not even describe the system qualitatively.
MPS and MPO are build to efficiently represent correlated states or operations. The first two top MPOs represent tight binding coefficients  $c^*_{\mathbf{k}_{i} n_1}$ and $c^*_{\mathbf{k}_{j} n_2}$ (e.g. with a maximal link dimension), but since both do not share a single index, the two MPOs are completely uncorrelated, so that a resulting MPO with acceptable precision will scale with D$^2$ in the link dimension.
On the other hand the third, middle MPO including the Coulomb potential $V_{\mathbf{q}}$ connects all incoming and outgoing indices with each other, therefore all bits are correlated within this part of the tensor network.
Therefore an efficient and accurate implementation starts with combining for example the third and the second MPO (cf. Fig. \ref{tntcoulomb}e)).
Then the result is merged with the fourth MPO to eliminate the tight binding index $n_2$ and exploits its correlation with the second MPO.  Subsequently the first and than fifth MPO is finally merged. 
In this way, we were able to get the best accuracy, since minimal links are required.

However even in this way, we could not achieve convergence within meV range, 
therefore the merged MPOs are only used for an initial calculation with DMRG. The final calculation with imaginary time propagation (ITP) avoids merging the stacked MPO for the Coulomb problem.

\section{Methods for obtaining eigenstates}
\begin{figure}[tb] 	
\centering
  \includegraphics[width=7.0cm]{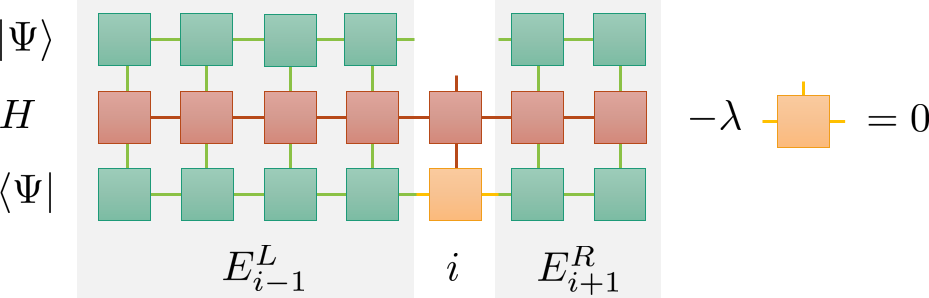}
  \caption{ Local eigenvalue problem solved at different site tensors $i$ during the sweeps. The size environment tensors $E^L_{i-1}$ and $E^R_{i+1}$ is crucial for the numerical effort of the DMRG.  }
  \label{dmrgfig}
\end{figure}
\subsection{Density matrix renormalization group (DMRG)}
DMRG is a method for obtaining the system ground state, first introduced in the seminal paper of S. White \cite{white1992density}.
The main objective is to find a MPS $|\Psi\rangle$, that minimizes 
$E=\langle \Psi |H |\Psi\rangle/\langle \Psi|\Psi \rangle$
according to Ritz's principle.
The problem is reformulated introducing the Lagrange multiplier $\lambda$ for the ground state and finding the extreme of \cite{schollwock2011density}
\begin{eqnarray}
\langle \Psi |H |\Psi\rangle-\lambda \langle \Psi |\Psi\rangle \label{DMRGminimize}
\end{eqnarray}
for retrieving the ground state $|\Psi\rangle$.
Instead of a variation of the full MPS $|\Psi\rangle$ only the tensor inside the MPS at a specific site $i$  is varied. By placing the orthogonality  center at site $i$, we retrieve a simplified eigenvalue problem to obtain the tensor, that minimizes Eq. (\ref{DMRGminimize}), (cf. Fig. \ref{dmrgfig} for illustration or \cite{schollwock2011density} for details). Important for the numerical efficiency are that the  tensors of the MPS and MPO, which are not at position $i$, are merged together into left and right environment tensor $E_{i-1}^L$ and $E_{i+1}^R$.
The link dimensions of the MPS and MPO at $i-1$ and $i+1$ determines the size of the environment tensors $E_{i-1}^L$ and $E_{i+1}^R$ and thus the numerical complexity of the problem. Therefore using stacked MPO for the Coulomb term will result in large environment tensors and thus a merging of the MPO is required.  

In DMRG
for obtaining the ground state MPS $|\Psi\rangle$, $i$ is several times swept from left to right and then from right to left, while performing at every position $i$ a minimization of the tensor $i$ inside the MPS (see for example the  review \cite{schollwock2011density} for details). 

DMRG is through modifications able to calculate excited states.
For example in the used itensor library \cite{itensor} projectors $|\psi_n\rangle\langle \psi_n |$  for already obtained eigenstates $|\psi_n\rangle$  are added into the minimization problem
\begin{eqnarray}
\langle \Psi |H |\Psi\rangle+ D \sum_n \langle \Psi |\psi_n\rangle \langle \psi_n |\Psi \rangle -\lambda \langle \Psi |\Psi\rangle \label{DMRGminimizeExcited}
\end{eqnarray}
together with a penalty energy $D$ (typical value for our simulations $20~ \mathrm{eV}$).
Other modified versions of DMRG for excited states were proposed in \cite{PhysRevB.94.045111,PhysRevLett.118.017201,PhysRevLett.116.247204}. 

We use the DMRG implementation of the itensor library (version 2.1.0 patched) \cite{itensor}, which is modified to use Hamiltonians represented as a sum of MPOs and to calculated excited states simultaneously.

\subsection{Imaginary time propagation(ITP)}
 While the DMRG method introduced in the last section is usually very fast and reliably allows to retrieve the ground state of most systems, we face major problems for the generalized Wannier equation. The main obstacles for the generalized Wannier equation are the tight-binding coefficients present in the Coulomb interaction (cf. tensor network in Fig. \ref{tntcoulomb}a)-c)), which requires a merging/multiplication of five MPOs (cf. Fig. \ref{tntcoulomb}e)). 
 Even with very low precision  very high link dimension in the order of thousands appear \cite{Kuhn:2019} during the multiplication. So that we could not achieve convergence for the electron-hole complexes studied here on our computing nodes.   Without including the tight binding coefficients (which may be justified for an Ansatz including only certain parts of the Brioullin zone), a calculation of larger electron hole complexes states is feasible using DMRG with calculation times around hours.
 Including the tight-binding coefficients the merging/multiplication of the involved MPO's consumes the majority of the time (cf. Fig. \ref{tntcoulomb}e)).
 
 Of course, achieving convergence is much easier, if MPO merging  is avoided. 
 So that instead of representing a single part of the Hamiltonian $H_{C}$ as one MPO, we keep the five unmerged MPO $H_{C}=H_{TB,1}H_{TB,2}H_{C,3} H_{TB,4} H_{TB,5}$.
 For DMRG, using the unmerged MPOs is not possible. In principle, the minimization of $\langle \Psi| H|\Psi\rangle-\lambda \langle \Psi|\Psi\rangle$ in DMRG could be written with unmerged MPOs  for the contribution $\langle \Psi| H_{C}|\Psi\rangle=\langle \Psi| H_{TB,1}H_{TB,2}H_{C,3} H_{TB,4} H_{TB,5}|\Psi\rangle$.
  However  the environment tensors ($E_{i-1}^L$ and $E_{i+1}^R$) inside the DMRG algorithm will scale roughly with $D^5 $.
  Typical values $D$ are in the range from 30-200 leading to $D^5$ in the range from $2.4\cdot 10^7$ to $3.2\cdot 10^{11}$, which is not feasible. This shows clear future  demand for a modification of the  DMRG algorithm.
 On the other hand, we can solve the time dependent Schrödinger equation:
 \begin{eqnarray}
 i\hbar\partial_t |\Psi\rangle= H |\Psi\rangle.
 \end{eqnarray}
 For calculating the time propagation, it is required to apply the Hamiltonian $H$ as MPO to the current MPS state $|\Psi\rangle$. For the unmerged MPO $H_{C}=H_{TB,1}H_{TB,2}H_{C,3} H_{TB,4} H_{TB,5}$, a successive application $|\chi^{(5)}\rangle=H_{TB,5} |\Psi\rangle$, $|\chi^{(4)}\rangle=H_{TB,4} |\chi^{(5)}\rangle$,..., $|\chi^{(1)}\rangle=H_{TB,2}$,  $|\chi^{(2)}\rangle=H_{C}|\Psi\rangle $ is possible and avoids the merging of MPO. The  MPS truncation $|\chi^{(i)}\rangle$ (the vector) can be achieved with much higher accuracy than  a MPO  (the matrix) truncation, since we hold only the information, how a single vector (MPS) is transformed by the Hamiltonian and not the information how any vector is transformed by the merged Hamiltonian (MPO).
 In practice, we use a variational approach \cite{schollwock2011density,itensor} for the application of the MPO to the MPS states, since the exact application is too slow. 
 
 A numerical solution of the  time dependent Schrödinger equation does not lead to the eigenvalues and eigenstates of the system. However, if the time $t$ is replaced by $t=-i\hbar\beta$, we arrive at an equation for imaginary time propagation (ITP) \cite{schollwock2011density}:
 \begin{eqnarray}
\partial_{\beta} |\Psi\rangle=-H|\Psi\rangle, \label{imag_time_prop}
 \end{eqnarray}
 with the formal solution $|\Psi\rangle(\beta)= e^{-\beta H}|\Psi\rangle(\beta=0)$.
  Expanding $|\Psi\rangle(\beta=0)$ into the eigenstates $|n\rangle$ of $H$ with eigenenergy $E_n$ yields $|\Psi\rangle(\beta=0)=\sum_n c_n |n\rangle$ with the formal $\beta$ dependent solution
\begin{eqnarray} 
|\Psi\rangle(\beta)=\sum_n c_n e^{-\beta E_n} |n\rangle.
\end{eqnarray}
 If $E_1<E_n$ holds for the ground state energy $E_1 $ and we define $|\tilde{\Psi}\rangle(\beta)=e^{-\beta E_1} |\Psi\rangle(\beta)$  we see immediately from
 \begin{eqnarray} 
 |\tilde{\Psi}\rangle(\beta)= |1\rangle +\sum_{n\neq 1} c_n e^{-\beta (E_n-E_1)} |n\rangle, 
 \end{eqnarray}
 that $ |\tilde{\Psi}\rangle(\infty)= |1\rangle $, i.e. for long times $ |\tilde{\Psi}\rangle$ propagates towards the ground state. Or more precisely the wavefunction propagates towards the eigenstate with the lowest energy included in the expansion $|\Psi\rangle(\beta=0)=\sum_n c_n |n\rangle$  with $c_n\neq 0$. This is exploited later for  the calculation of excited states.
 Usually propagation starts with a random MPS for $|\Psi\rangle(\beta=0)$, which should include contributions from the lowest energy eigenstate. Numerical errors or errors due to the MPS approximation are  beneficial for the algorithm, since these may expand the number of eigenstates included in $|\Psi\rangle(\beta)$. 
 
 For the implementation we take Eq. (\ref{imag_time_prop}) and replace $H$ with $H+\lambda Id$ using $\lambda$ as parameter to shift the energy axis and use a first order Euler approximation:
 \begin{eqnarray}
 |\Psi(\beta+\Delta\beta)\rangle=(Id-\Delta\beta(H+\lambda Id)) |\Psi(\beta)\rangle
 \end{eqnarray}
 for calculating the ITP using a variational algorithm  algorithm \cite{schollwock2011density,itensor} (fitMPOapply from itensor), than $|\Psi(\beta+\Delta\beta)\rangle$ is immediately normalized after every time step.
 Before normalization $\langle \Psi(\beta)|H|\Psi(\beta)\rangle
 =1-\langle \Psi(\beta)|\Psi(\beta+\Delta\beta)\rangle/\Delta\beta$ allows to calculate the average energy.
 Furthermore  $\langle \Psi(\beta+\Delta\beta)| \Psi(\beta+\Delta\beta)\rangle=  \langle\Psi(\beta)| (Id-\Delta\beta(H+\lambda Id))^2 |\Psi(\beta)\rangle$ and recognizing that $\langle \Delta H\rangle = \langle \Delta (Id-\Delta\beta(H+\lambda Id)) \rangle$ allows to decide if the propagation is close to a single eigenstate, since for a single eigenstate the propagation reaches $\langle \Delta H\rangle=0$, if $\langle \Delta H\rangle$ is calculated without approximations.

\subsection{Combined approaches}
In practice, it is useful to start the calculation of an eigenstate with a DMRG calculation. Even if the result is not converged, the overall structure of the eigenstates is in general close enough to the converged result, that propagation until convergence in a subsequent ITP is drastically  faster using the  DMRG result  as starting point for the propagation. 
This works always for the ground state inside the subspace of the calculation, since ITP will always lead to the ground state. 
However some times the variational application of the MPO  in the ITP   is stuck inside a subspace, in these cases a subsequent ITP starting from the excited states from DMRG will also work for retrieving excited states. We will discuss this in detail for the different applications (see section \ref{convergentexcited}). 

\section{Exciton states}
\label{exciton_section}
\begin{figure}[tb] 	
\centering
  \includegraphics[width=7.0cm]{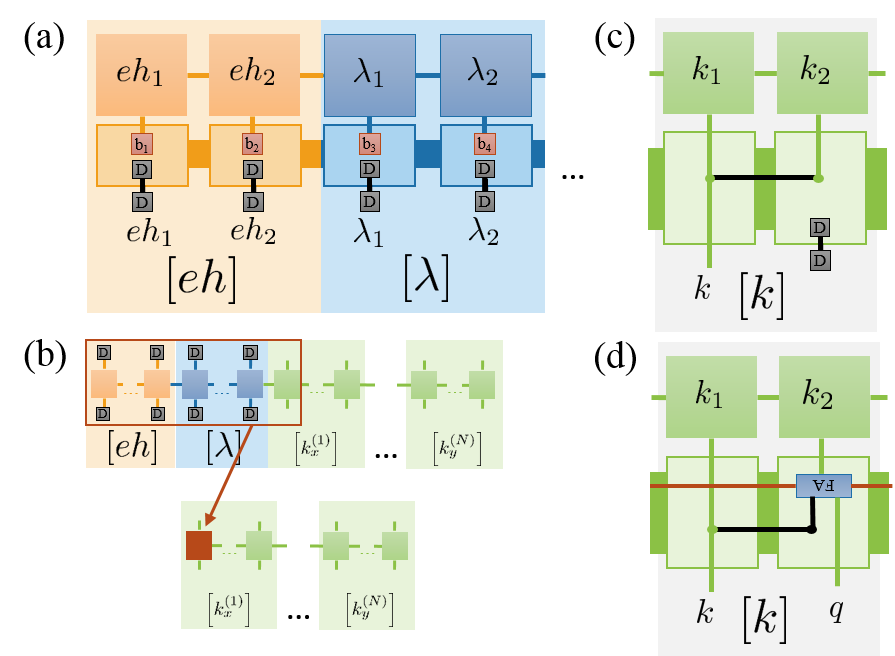}
  \caption{ a) Combined projector/transformation $PU$ for setting the band $eh$ and spin index $\lambda$ bits to the bits $b_1$.. $b_4$, b) after the $PU$ and $U^\dagger P$ is applied to the MPO, the unnecessary bits are terminated by tensor $D$ and the corresponding tensors of the MPO are joined with the next remaining tensor, c) projector/transformation part for one $\mathbf{k}$ bit for bright excitons and d) transformation for setting $\mathbf{k}_1=\mathbf{k}$ and $\mathbf{k}_2=\mathbf{k}+\mathbf{q}$.  }
  \label{excitonproj}
\end{figure}
\subsection{Coordinate transformations}
In most treatments only the optical active exciton states are of interest, this is routinely done \cite{haug2009quantum,berghauser2014analytical,mostaani2017diffusion,schmidt2016ultrafast,richter2017nanoplatelets,C9NR03161H,PhysRevLett.111.216805} without tensor networks. 
For the tensor network method the optical active states are a benchmark and to illustrate and test concepts, which are applied for larger electron hole complexes (trion and biexcitons). Exciton states in general benefit from the tensor networks, since higher grid sizes are easily possible as well as test of  common approximation and physical mechanism such as a restriction to a part of the Brioullin zone, hybridization between K and K' excitons and effective mass description. 

Starting point for  the exciton wavefunction
is the two particle wavefunction $\Psi_{\mathbf{k}_1 \mathbf{k}_2}$ together with the generalized Bethe-Salpeter Eq. (\ref{bethesalpeterlike}).
However the calculation of the ground state for $\Psi_{\mathbf{k}_1 \mathbf{k}_2}$ will not yield an exciton but the wave function  of two holes, since so far all possible combinations of bands, spins are included in the MPO implementation of Eq. (\ref{bethesalpeterlike}).
So for retrieving an exciton $\Psi_{h\lambda_1\mathbf{k}_1 e\lambda_2\mathbf{k}_2}$ is calculated, setting for example the first particle to conduction band $e$ and spin $\lambda_1$ and the second to valence band $h$ and spin $\lambda_2$. 
Furthermore often a calculation of optical active excitons is desired with only diagonal contributions to the exciton wave function  $\Psi_{h\lambda_1\mathbf{k} e\lambda_2\mathbf{k}}\neq0$.
The restriction to a certain combination of bands and spins as well as taking only diagonal optical active states are both projections $P$ into a subspace of the Hilbert space.
We will use in the following a combination $PU$ of a projection $P$ and a coordination transformation $U$. 
For applications together with the tensor networks, $PU$ are MPOs and are presented here for the different index transformations (including the other quasi particles).

Every MPO $H_{MPO}$ describing a part of the term in the Eq. (\ref{bethesalpeterlike}) is replaced with  $ P U H_{MPO}  U^\dagger P$ in the calculation (see Fig. \ref{tntcoulomb}f)(i)). 
$P$ has to be chosen, so that every $H_{MPO}$ is diagonal in the subspace, i.e.  $P$, $P U H_{MPO} U^\dagger (Id-P)=(Id-P)U H_{MPO}U^\dagger P=0$ holds. This requirement also applies for every of the stacked MPOs for the Coulomb term (cf. Fig. \ref{tntcoulomb}f)(ii)), since $PU$ is applied to each MPO individually.
It is also imperative to remove sites from the MPSs and MPOs, so that only the subspace of $P$ is indexed by them.
Otherwise for example DMRG will calculate a state inside $Id-P$, for which the eigenvalue will be $0$, since $P U H U^\dagger P  (Id-P)=0$. 

We  illustrate the construction of a projector $PU$ for the bits for band and spin, i.e. for $\Psi_{h\lambda_1\mathbf{k}_1 e\lambda_2\mathbf{k}_2}$, in Fig. \ref{excitonproj}a). The projector terminates the bits with the selected band or spin and the site dimension is  set to $0$ using a tensor $D$. 
The projector is than applied to the original MPO $H$ as $PUHU^\dagger P$, see Fig. \ref{tntcoulomb}f). In a second step, the tensor of the MPO with fixed site bits are terminated by a tensor $D$ and merged with the adjacent tensors inside the MPO (see Fig. \ref{excitonproj}b)). 

A projector restricting the calculation to optical active excitons with diagonal index, e.g. $\Psi_{h\lambda_1\mathbf{k} e\lambda_2\mathbf{k}}$, is constructed by connecting the two adjacent $k$ site bits to a delta tensor as depicted in Fig. \ref{excitonproj}c). One of the bit is obsolete after connecting the delta tensor and the connected tensor is removed after applying the projector $P$ to the Hamiltonian $UPHPU^\dagger$ by merging  with an adjacent tensor (analog to the band and spin indices cf. Fig. \ref{excitonproj} a) and b)).

\begin{figure}[tb] 	
\centering
\includegraphics[width=5.0cm]{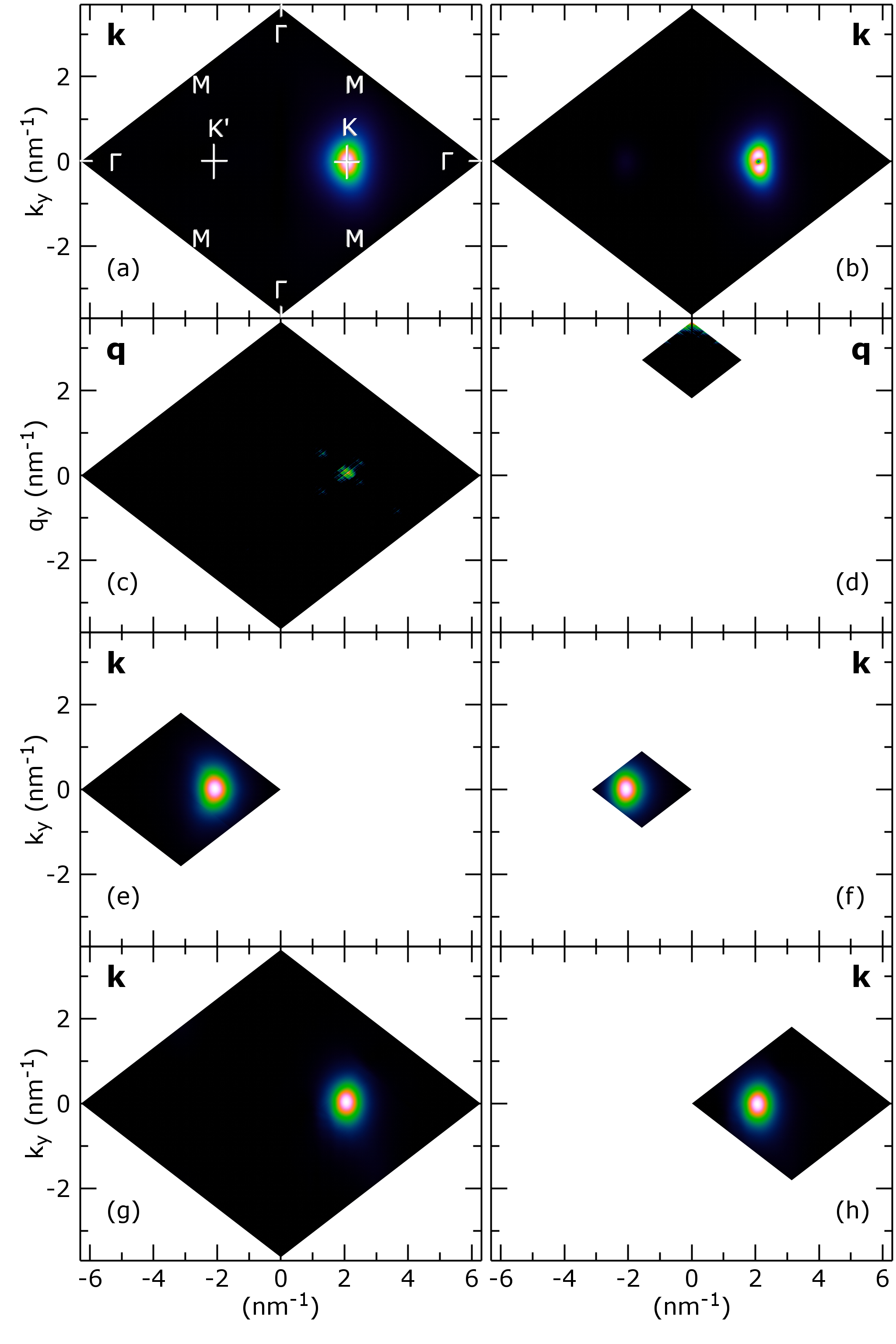}
\caption{Selected exciton wave functions in BZ:
   (a) bright A 1s and (b) A 2p exciton $\Psi_{h\uparrow\mathbf{k} e\uparrow\mathbf{k}}$ with parallel spin up,
   $\mathbf{q}$ distribution (logarithmic) $\sum_{\mathbf{k}}\Psi_{h\uparrow\mathbf{k} e\uparrow\mathbf{k}+\mathbf{q}}$ for (c) the momentum forbidden 1s exciton and (d) the bright 1s A exciton with $\mathbf{q}$ restricted to a box,
   B exciton distribution $\sum_{\mathbf{q}}\Psi_{h\uparrow\mathbf{k} e\uparrow\mathbf{k}+\mathbf{q}}$ calculated using  (e) a big box or (f) small box to confine $\mathbf{k}$.
   The outside of boxes is plotted as white.
   KQ (g) exciton wave function  $\Psi_{h\uparrow\mathbf{k} e\uparrow\mathbf{k}+\mathbf{q}}$ calculated with fixed $\mathbf{q}$ (h) exciton distribution $\sum_{\mathbf{q}}\Psi_{h\uparrow\mathbf{k} e\uparrow\mathbf{k}+\mathbf{q}}$ with $\mathbf{k}$ and $\mathbf{q}$ confined to a box.
         }
  \label{excitonfig}
\end{figure}

\begin{figure}[tb] 	
\centering
  \includegraphics[width=7.0cm]{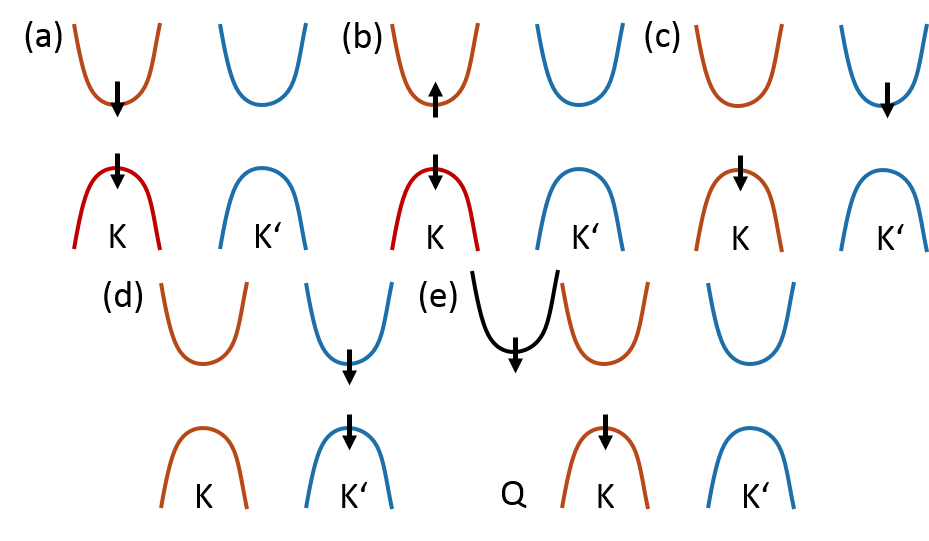}
  \caption{  Overview of the example electron hole configuration of the  calculated exciton states.  }
  \label{excitonconfig}
\end{figure}

\begin{table}[]
\begin{tabular}{|l|l|l|l|l|}
\hline
 State & DMRG  & ITP            & Restriction                                     & Fig.\ref{excitonconfig} \\
 \hline
 ground state & {\it 1.762  }            & 1.767          & none                                            &   c)  \\

     {\footnotesize(momentum dark)}                & {\it 1.764 }             & 1.769          & big box $\mathbf{k}$ at K                       &        \\
   
                     & {\it 1.785 }             & 1.790          & small box $\mathbf{k}$ at K                     &        \\
                      \hline
  A 1s (bright)        & {\it 1.758  }            & \textbf{1.769} & $\mathbf{k}_1=\mathbf{k}_2$                     & a)     \\
                     & {\it 1.766 }             & \textbf{1.769} & box  $\mathbf{q}\approx 0$                      &        \\
 (spin dark)     & {\it 1.755 }             & 1.763          & $\mathbf{k}_1=\mathbf{k}_2$                     & b)     \\
 \hline
B 1s (bright)        & {\it 1.897 }             & 1.906          & $\mathbf{k}_1=\mathbf{k}_2$                     & d)     \\
                     & {\it 1.904 }             & 1.911          & big box $\mathbf{k}$ at K'                      &        \\
                     & {\it 1.927 }             & 1.934          & small box $\mathbf{k}$ at K'                    &        \\
(spin dark)     & {\it 1.900 }             & 1.912          & $\mathbf{k}_1=\mathbf{k}_2$                     &        \\
\hline
A 2p (bright)        &                    & 1.971          & $\mathbf{k}_1=\mathbf{k}_2$                     &        \\
 (spin dark)     &                    & 1.969          & $\mathbf{k}_1=\mathbf{k}_2$                     &        \\
 \hline
B 2p (bright)        & {\it 1.802 }             &                & $\mathbf{k}_1=\mathbf{k}_2$                     &        \\
 (spin dark)     & {\it 1.802 }             &                & $\mathbf{k}_1=\mathbf{k}_2$                     &        \\
 \hline
K-Q                  & {\it 1.950 }             & 1.936          & $\mathbf{q}$ (i) fixed value                        & e)     \\
                     & {\it 1.946 }             & 1.938          & (ii)fix upper bits of  & \\
                     & & &$\mathbf{k}$ and $\mathbf{q}$ & \\
\hline      
\end{tabular}
 \caption{ \SANchange{ Numeric results for exciton energies (in eV) comparing different strategies of the calculation. DMRG results did not achieve convergence (italics).} }
 \label{exciton_table}
\end{table}

\subsection{Numerical results}
\label{exciton_num_res}
We start with a discussion of numerical results for the diagonal bright exciton $\Psi_{h\lambda_1\mathbf{k} e\lambda_2\mathbf{k}}$.
\SANchange{
For example a DMRG calculation, chosen so that it is a fast and good starting point for ITP but not necessarily with accurate energies, yields for the lowest three bright and (spin forbidden dark) A 1s exciton (Fig. \ref{excitonconfig}a)), B 2p exciton, B 1s exciton (Fig. \ref{excitonconfig}d)) etc. the energies as given in Table \ref{exciton_table} (with restriction $\mathbf{k}_1=\mathbf{k}_2$).}
As written earlier, we could not achieve convergence for the DMRG case due to the tight binding coefficients.
Hence all DMRG calculations in this paper provide initial states for a fast ITP calculation and are calculated with reduced accuracy \SANchange{(cf. in Table \ref{exciton_table} the unusual energy ordering of the states)}.

\SANchange{
A calculation using ITP, which numerical converged, yields very different results for the bright and dark  A 1s exciton (Fig. \ref{excitonconfig}a)), B 1s exciton (Fig. \ref{excitonconfig}d)), A 2p exciton  and (cf. Table \ref{exciton_table}).}
Taking the DMRG result as start point for the ITP yields the same energy of $1.769~\mathrm{eV}$ for the A 1s exciton.
 For the excited states the procedure does not work, since in this case the ITP always retrieves the ground state, the A 1s exciton. 

In order to get reliable results, a careful convergence analysis is required for the MPS of the band structure, Coulomb potential etc. as well as the parameters of DMRG or ITP \REVchange{for excitons, but also for trions and biexcitons}, see supplemental material of Ref. \onlinecite{Kuhn:2019} \REVchange{and also Sec. \ref{convergentexcited}} for the analysis and parameters used here.
Indeed the large grid size of $1024\times 1024$ is necessary for a high accuracy of the exciton binding energy. \REVchange{ The convergence analysis also ensures, that we do not encounter problems from a potential unfavorable entropy scaling in two dimensions\cite{RevModPhys.82.277}. }

Calculating the bright $k$-diagonal exciton states is not special and routinely carried out without tensor networks \cite{haug2009quantum,berghauser2014analytical,mostaani2017diffusion,schmidt2016ultrafast,richter2017nanoplatelets,C9NR03161H,PhysRevLett.111.216805}.
However tensor networks enable also to calculate $\Psi_{h\lambda_1\mathbf{k}_1 e\lambda_2\mathbf{k}_2}$ without setting $\mathbf{k}_1=\mathbf{k}_2$, so that deviations from a center of mass motion and momentum forbidden excitons  are addressed. 
The index transformation $\Psi_{h\lambda_1\mathbf{k} e\lambda_2\mathbf{k}+\mathbf{q}}$ is beneficial, since Coulomb interaction will leave $\mathbf{q}$ invariant and thus $\mathbf{q}$ should be a delta peak at a single $\mathbf{q}$.
Furthermore similar index transformations are required for trions and biexcitons, so that concepts are tested here at the known exciton level.
The index transformation $PU$ of $\mathbf{k}_1$ and $\mathbf{k}_2$ to $\mathbf{k}$ and $\mathbf{k}+\mathbf{q}$ is handled by a fulladder circuit similar to the Coulomb interaction (Fig. \ref{excitonproj}d)).
The DMRG calculation as input for ITP reveals an ground state exciton energy of 1.762 eV 
and a subsequent ITP yields 1.767 eV.
The energies are different to the previous calculation using diagonal $\mathbf{k}$ momenta \SANchange{($\mathbf{k}_1=\mathbf{k}_2$, cf. Table \ref{exciton_table})}.
For the seed DMRG deviations are expected, but not for ITP.
However, an inspection of the exciton wave function shows that here the ground state is an optically momentum forbidden
K-K' exciton (see Fig. \ref{excitonfig}c)) with electron and holes being at different symmetry points. 
Therefore the desired bright 1s A exciton is an excited state in this case an ideal testing case for strategies needed later for the bright biexciton.    

The excited states retrieved by DMRG are completely unordered but the low energy states are also momentum forbidden 1s K-K' excitons and are describing the center of mass continuum of the ground state.
In this way the $1024\times 1024$ grid used for the Brillouin Zone is a curse (it also haunts the biexciton problem), since  a huge number of dark continuum states for the ground state  is retrieved more than  is actually feasible to also access other excited states.
Therefore strategies for calculating selectively the excited states are required. 

One way is to set $\mathbf{q}$ to a fixed value (the optical allowed diagonal case is the special case $\mathbf{q}=0$ ), which is exact, since the Coulomb interaction leaves $\mathbf{q}$ invariant. In this case, the $\mathbf{q}$ bits  are set to the fixed value and removed afterwards, analog to the procedure for the band and spin indices.
 Often the exciton state with the smallest energy with electron and holes at certain symmetry points is desired or the momentum dark state instead of the bright state or vice versa. In this case, we  set the upper bits of $\mathbf{q}$ to a fixed value, restricting $\mathbf{q}$ to a box around bit boundaries (cf. Fig. \ref{excitonfig}d)). 
 
 On the other hand the same can be applied to  $\mathbf{k}$, e.g. to focus only on certain symmetry points.
 If the boxes do not fit well to the desired symmetry points, the Brillouin zone can be shifted by $k_x^{shift}$ and $k_y^{shift}$. Restricting $\mathbf{k}$ is actually an approximation, since superpositions between different symmetry points due to Coulomb interaction are prohibited. However a restriction to certain symmetry points is a common approximation \cite{berghauser2014analytical,steinhoff2018biexciton}, which we test here:
The bright A exciton is calculated
by restricting the momentum $\mathbf{q}$ to a box including $\mathbf{q}\approx 0$, therefore fixing the upper two bits of every dimension to 0 (cf. Fig. \ref{excitonfig}d)).
An energy of 1.766 eV
is retrieved in the seed DMRG and of 1.769 eV in the subsequent ITP,
showing no deviation from the calculation of the bright exciton before in ITP \SANchange{(cf. Table \ref{exciton_table})}.

For retrieving higher excited states such as the bright B exciton
we fix the upper bits of $\mathbf{k}$ to restrict it to a box around the symmetry point  such as K or K'.
For the momentum forbidden ground state exciton (Fig. \ref{excitonconfig}c)), we retrieve 1.764 eV (1.785 eV) in DMRG and
 1.769 eV  (1.790 eV) in ITP for a big (small) box (cf. Fig. \ref{excitonfig}c)).
For the 1s B exciton  the energy is 1.904 eV (1.927 eV) in DMRG
 and 1.911 eV (1.934 eV) in ITP for a big (small) box (cf. Fig. \ref{excitonfig}e)f)).
The approach does not include couplings/hybridization between the K-K' point (cf. \cite{berghauser2018inverted}) and thus changes the energies slightly by a few $\mathrm{meV}$,  showing that for MoS$_2$ the hybridization and thus the Coulomb hybridization (cf. "Dexter" coupling in \cite{berghauser2018inverted}) leads to a shift of several meV and that the full Brioullin zone should be included in the calculation. 
Also it is clear, that the box has to be big enough to include the full wave function (cf. the error of over 10 meV for a small box).
  
Furthermore K-Q excitons are in the focus of current investigations \cite{maltedarkbright,deilmann2019finite}. 
Two strategies are possible to access these:
(i) $\mathbf{q}$ is set to a fixed value (the distance between K-Q) and (ii)  the upper bits of $\mathbf{q}$ and $\mathbf{k}$ are fixed, so that the state is found inside these boxes. \SANchange{The results for both cases are given in Table \ref{exciton_table}.}
\SANchange{ We notice} a small shift of $\mathbf{q}$ in case of (ii) compared to the distance between K and Q, which was used in case (i) for the fixed $\mathbf{q}$. The difference of a few eV \SANchange{(cf. Table \ref{exciton_table})} may be solely attributed to the box used to confine $\mathbf{k}$.
Note, that in case of fixed $\mathbf{q}$, we can easily access excited states like 2s, 2p etc, since it is similar to the diagonal bright case.

\section{Trion states}
\label{trion_section}
\begin{figure}[tb] 	
\centering
  \includegraphics[width=8.0cm]{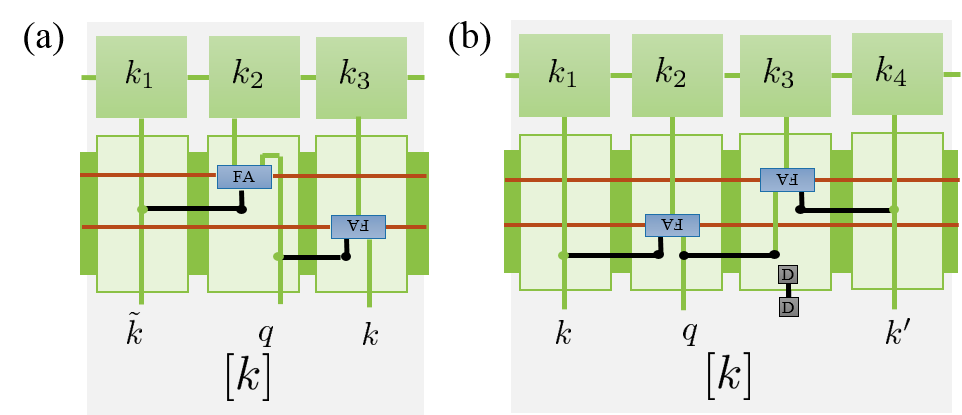}
  \caption{  (a) Transformation for trions setting $\mathbf{k}_1=\tilde{\mathbf{k}}$, $\mathbf{k}_2=\tilde{\mathbf{k}}-\mathbf{q}$ and $\mathbf{k}_3=\mathbf{k}+\mathbf{q}$  and (b) projection/transformation for biexcitons setting $\mathbf{k}_1=\mathbf{k}$, $\mathbf{k}_2=\mathbf{k}+\mathbf{q}$, $\mathbf{k}_3=\mathbf{k}'+\mathbf{q}$ and $\mathbf{k}_4=\mathbf{k}'$.  }
  \label{trionbiexcproj}
\end{figure}
The trion is the next bigger bound electron hole complex 
after the exciton. A trion is a charged exciton with an additional electron or hole included in the complex.
In the low density limit, the optical created trions does not start from the neutral ground state $|\psi_0\rangle$, but from a state with already an hole $a^\dagger_{h\lambda\mathbf{k}} |\psi_0\rangle$ or an electron $a^\dagger_{e\lambda\mathbf{k}} |\psi_0\rangle$ with momentum $\mathbf{k}$ present in the system, corresponding to the wavefunctions $\Psi_{h\lambda_1\mathbf{k}}$ or $\Psi_{e\lambda_1\mathbf{k}}$. The momentum added by optical excitation of the additional electron hole pair is negligible, thus that the overall momentum of a bright trion is $\mathbf{k}$ the momentum of the initial carrier.
In general the wavefunction of a trion is for T$^-$: $\Psi_{h\lambda_1\mathbf{k}_1 e \lambda_2\mathbf{k}_2 e \lambda_3\mathbf{k}_3}$ and T$^+$: $\Psi_{e\lambda_1\mathbf{k}_1 h \lambda_2\mathbf{k}_2 h \lambda_3\mathbf{k}_3}$, for which the generalized Bethe Salpeter Eq. (\ref{bethesalpeterlike}) holds.
If only bright trion complexes are of interest, the 
coordinate transformation $\Psi_{h\lambda_1\tilde{\mathbf{k}} e \lambda_2 \tilde{\mathbf{k}}-\mathbf{q}  e \lambda_3\mathbf{k}+\mathbf{q}}$ or $\Psi_{e\lambda_1\tilde{\mathbf{k}} h \lambda_2 \tilde{\mathbf{k}}-\mathbf{q}  h \lambda_3\mathbf{k}+\mathbf{q}}$   is beneficial with momentum $\mathbf{k}$ of the initial carrier or the whole trion.  
Since the Coulomb interaction does not change $\mathbf{k}$ in these coordinates, $\mathbf{k}$ can be set to a fixed value.
The coordinate transformation UP is implemented as MPO including two fulladder circuits for every k bit (see Fig. \ref{trionbiexcproj}a)).
For the calculation the k bits can be fixed analog to fixing q bits in the exciton case.
A separate calculation of the trion eigenstates is required  for every $\mathbf{k}$.
The calculation is very efficient, but we can not calculate trion eigenstates for the whole 1024x1024 grid.
If trion binding energies are of interest, the carrier momentum $\mathbf{k}$ is placed at band minima at typical symmetry points. If a calculation of the line shape is of interest as in Ref. \onlinecite{zimmermanntrion2000}, a coarse grid around the symmetry points can be used to retrieve an approximation of the overall shape.

As in the exciton case, the trion states are first calculated with a low accuracy DMRG calculation and a subsequent ITP achieves convergence. While in the exciton case the link dimension of the MPS during ITP was effectively unrestricted, we had to restrict the link dimension in the trion and biexciton case. 
For a fast calculation the ITP typically starts with a restricted link dimension of 200, after being close to the final state the ITP continues briefly with a link dimension of 500. (See biexciton section \ref{convergentexcited} for a detailed analysis of the strategy.)

\begin{figure}[tb] 	
\centering
  \includegraphics[width=8.0cm]{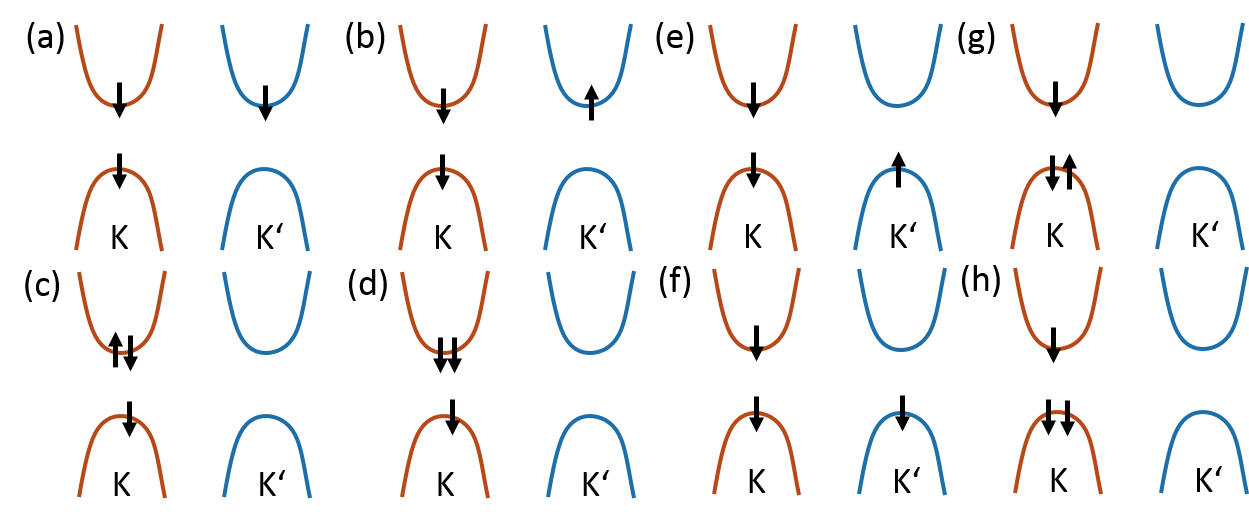}
  \caption{ \SANchange{ Overview of the example electron hole configuration of the  calculated trion states.}  }
  \label{trionbiexcschemes}
\end{figure}
\begin{table}[]
\begin{tabular}{|l|l|l|l|l|l|l|}
\hline
 Type & Configuration  & $E_t$ & $E_b$ & $\mathbf{k}$            & Restriction &Fig. \ref{trionbiexcschemes}                                   \\
 \hline
 T$^-$ A &$\Psi_{h\downarrow\tilde{\mathbf{k}} e \downarrow \tilde{\mathbf{k}}-\mathbf{q}  e \downarrow \mathbf{k}+\mathbf{q}}$
 &1.742 & 27 & K' & none & a) \\
 &  & 1.746 & 23 & K & & d)\\
 T$^-$ B &  & 1.885 & 21 & K & Sec.\ref{convergentexcited} & \\
  T$^-$ A & $\Psi_{h\downarrow\tilde{\mathbf{k}} e \uparrow \tilde{\mathbf{k}}-\mathbf{q}  e \downarrow \mathbf{k}+\mathbf{q}}$ & 1.756 & 13 & K' & $\mathbf{q}$ box K'& b) \\
  & & 1.748  & 21 &K'& big $\mathbf{q}$ box K'& \\
  &$\Psi_{h\downarrow\tilde{\mathbf{k}} e \downarrow \tilde{\mathbf{k}}-\mathbf{q}  e \uparrow \mathbf{k}+\mathbf{q}}$&
  1.7425 & 27 & K & none & c)\\
  \hline
  T$^+$ A & $\Psi_{e\downarrow\tilde{\mathbf{k}} h \downarrow \tilde{\mathbf{k}}-\mathbf{q}  h \uparrow\mathbf{k}+\mathbf{q}}$ & 1.740 & 29 & K'& none & e)\\
  & $\Psi_{e\downarrow\tilde{\mathbf{k}} h \downarrow \tilde{\mathbf{k}}-\mathbf{q}  h \downarrow\mathbf{k}+\mathbf{q}}$&
  1.736 & 33 &K ' & none & f)\\
  & & 1.739 &29.4 & K &  & h)\\
  & $\Psi_{e\downarrow\tilde{\mathbf{k}} h \downarrow \tilde{\mathbf{k}}-\mathbf{q}  h \downarrow\mathbf{k}+\mathbf{q}}$
  &1.739 & 30 & K & box $\mathbf{k}$ & g) \\ \hline
\end{tabular}
 \caption{ \SANchange{ Numeric results for trion transition energies $E_t$ (in eV) and binding energies $E_b$ (in meV) from ITP    comparing different strategies of the calculation. } }
 \label{trion_table}
\end{table}
\begin{figure}[tb] 	
\centering
  \includegraphics[width=4.0cm]{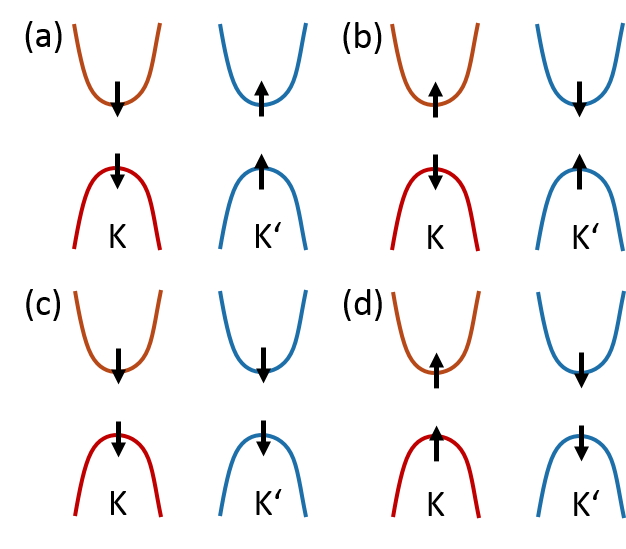}
  \caption{  Overview of the example electron hole configuration of the  calculated biexciton states.  }
  \label{biexcitonconfig}
\end{figure}

We start with example states for the negatively charged trion T$^-$ $\Psi_{h\downarrow\tilde{\mathbf{k}} e \downarrow \tilde{\mathbf{k}}-\mathbf{q}  e \downarrow \mathbf{k}+\mathbf{q}}$ \SANchange{with $\mathbf{k}$ at the K' point. The resulting trion has  an electron hole pair at the K point and an excess electron at K' as depicted in Fig. \ref{trionbiexcschemes} a) with a transition energy of 1.742 eV and  a trion binding energy of 27 meV. }
The transition energy \SANchange{$E_t$} of the trion is the difference between trion energy and energy of initial free carrier with momentum $\mathbf{k}$.
The binding energy \SANchange{$E_b$} is the difference between the transition energy and the exciton energy.
\SANchange{
For getting the configuration in Fig. \ref{trionbiexcschemes}b), we use an initial electron at K' with spin up (flipped compared to a)) and restrict $\mathbf{q}$ to a box around K' (cf. section \ref{exciton_num_res}) for the wavefunction $\Psi_{h\downarrow\tilde{\mathbf{k}} e \uparrow \tilde{\mathbf{k}}-\mathbf{q}  e \downarrow \mathbf{k}+\mathbf{q}}$.
 A small box (upper two bits fixed)  yields a transition energy of 1.756 eV in ITP and a binding energy of 13 meV.
A bigger box (upper bit fixed) results in a transition energy of 1.748 eV in ITP and a binding energy of 21 meV.
So in contrast to the exciton case, where $\mathbf{q}$ is delta-like, the introduction of a too small box introduces a substantial error for the broad  $\mathbf{q}$ distribution for the trion. A small error can remain for the bigger box (cf. Sec. \ref{biexciton_section}).}

For an initial carrier $\mathbf{k}$ with $\uparrow$ at the K point, we yield for the spin configuration $\Psi_{h\downarrow\tilde{\mathbf{k}} e \downarrow \tilde{\mathbf{k}}-\mathbf{q}  e \uparrow \mathbf{k}+\mathbf{q}}$ a trion T$^-$ with all carriers at the K point as depicted in Fig. \ref{trionbiexcschemes} c), ITP results in a transition energy of 1.7425 eV and a binding energy of 27 meV.
So only a slightly higher transition energy as the configuration with equal spin and the K'  point electron.
On the other hand also a T$^-$ configuration $\Psi_{h\downarrow\tilde{\mathbf{k}} e \downarrow \tilde{\mathbf{k}}-\mathbf{q}  e \downarrow \mathbf{k}+\mathbf{q}}$ with equal spin with all electron and holes at K point  (initial electron also at K point) exist as shown in Fig. \ref{trionbiexcschemes}d) with higher transition energy 1.746 eV and lower 23 meV binding energy.  
Note, often this state is excluded using Pauli exclusion principle as argument, but this holds only for level systems like quantum dots, the existence of the triplet states is well known for quantum wells \cite{PEETERS2001139,PhysRevB.54.R2296} and the reduction of the binding energy is a result of the interplay of Pauli exclusion principle and Hund's rule.
From calculating $\Psi_{h\downarrow\tilde{\mathbf{k}} e \downarrow \tilde{\mathbf{k}}-\mathbf{q}  e \downarrow\mathbf{k}+\mathbf{q}}$ we also retrieved an excited $T^-$ containing a B exciton using the technique described in section \ref{convergentexcited} and an initial maximum link dimension 100 for ITP (in principle  Fig. \ref{trionbiexcschemes} a) with K and K' exchanged) with transition energy 1.885 eV and binding energy 21 meV.

We continue with example states for the positive charged trion T$^+$ for the spin configuration $\Psi_{e\downarrow\tilde{\mathbf{k}} h \downarrow \tilde{\mathbf{k}}-\mathbf{q}  h \uparrow\mathbf{k}+\mathbf{q}}$ with $\mathbf{k}$ at the K' point, we yield the trion as depicted in Fig. \ref{trionbiexcschemes} e) with $\uparrow$ at the K' point. ITP gives a transition energy of 1.740 eV and a binding energy of 29 meV.
For the equal spin configuration  $\Psi_{e\downarrow\tilde{\mathbf{k}} h \downarrow \tilde{\mathbf{k}}-\mathbf{q}  h \downarrow\mathbf{k}+\mathbf{q}}$ also with   $\mathbf{k}$ at the K' point (cf. Fig. \ref{trionbiexcschemes} f)), ITP results in a transition energy of 1.736 eV and a binding energy of 33 meV. Here the different spin of the excess carrier slightly alters the Coulomb contribution. 
\SANchange{
For T$^+$ also trions with $\mathbf{k}$ at the K point exist, however for retrieving the state  using the configuration $\Psi_{e\downarrow\tilde{\mathbf{k}} h \uparrow \tilde{\mathbf{k}}-\mathbf{q}  h \downarrow\mathbf{k}+\mathbf{q}}$, $\mathbf{k}$ had to be restricted around K (cf. Fig. \ref{trionbiexcschemes} g)) and yield after ITP a transition energy of 1.739 eV and a binding energy of 30 meV compared to the bright exciton.
So compared to the case with $\mathbf{k}$ at the K'-point,
the T$^+$ with equal spins $\Psi_{e\downarrow\tilde{\mathbf{k}} h \downarrow \tilde{\mathbf{k}}-\mathbf{q}  h \downarrow\mathbf{k}+\mathbf{q}}$ and with $\mathbf{k}$ at the K point does not require a restriction of $\mathbf{k}$ and yield a transition energy of 1.739 eV and a binding energy of 29.4 meV, so only slightly changed compared to the different spin case (cf. Fig. \ref{trionbiexcschemes} h)). 

Furthermore the differences between having all carriers at the same or at different $K$ or $K'$ are not so pronounced for the positive as for the negative charged case. This might be a consequence of lower effective masses. }

Overall the order of magnitude of the calculated trion binding energies agree  with  current literature\cite{Drueppel2017,PhysRevB.95.081301,PhysRevB.92.205418}.
For the trions one drawback of the method was, that we had to guess the spin configuration, the position of the initial carrier and also to use constraints for some states. So we had to select beforehand the configuration of interest and did not obtain a complete set of states, therefore only selected examples are given. A systematic study of the states is subject to future work.

\begin{table}[]
\begin{tabular}{|l|l|l|l|l|l|l|}
\hline
 Type & Configuration  & $E$ & $E_b$ &  Method &Fig.~\ref{biexcitonconfig}                                   \\
 \hline
 A-A (dark) &  $\Psi_{h\downarrow\mathbf{k} e \downarrow\mathbf{k}+\mathbf{q} h \uparrow\mathbf{k}'+\mathbf{q} e \uparrow\mathbf{k}' }$ &3.506 & 27 & Sec. \ref{convergentexcited} & b) \\ \hline
 A-A  & $\Psi_{h\downarrow\mathbf{k} e \downarrow\mathbf{k}+\mathbf{q} h \uparrow\mathbf{k}'+\mathbf{q} e \uparrow\mathbf{k}' }$ & 3.519 & {\tiny 19-20} & Sec. \ref{convergentexcited} & a) \\
 (bright)&   $\Psi_{h\uparrow\mathbf{k} e \downarrow\mathbf{k}+\mathbf{q} h \downarrow\mathbf{k}'+\mathbf{q} e \uparrow\mathbf{k}' }$ 
  & 3.536 & 1-2 & small $\mathbf{q}$ & \\
 & & 3.523 & 15 & big  $\mathbf{q}$ & \\
    & &   && box at K' & \\
 & & 3.518 & 20  &   Sec. \ref{convergentexcited} &\\ \hline  
 B-B &  $\Psi_{h\uparrow\mathbf{k} e \downarrow\mathbf{k}+\mathbf{q} h \downarrow\mathbf{k}'+\mathbf{q} e \uparrow\mathbf{k}' }$ & 3.791 & 21 & Sec. \ref{convergentexcited} & d) \\ \hline
 A-B & $\Psi_{h\downarrow\mathbf{k} e \downarrow\mathbf{k}+\mathbf{q} h \downarrow\mathbf{k}'+\mathbf{q} e \downarrow\mathbf{k}' }$ & 3.655 & 20 &  Sec. \ref{convergentexcited} & c) 
 \\ \hline
\end{tabular}
 \caption{ \SANchange{ Numeric results for 1s biexciton energies $E$ (in eV) and binding energies $E_b$ (in meV) from ITP  comparing different calculation strategies. } }
 \label{biexc_table}
\end{table}

\begin{figure}[tb] 	
\centering
\includegraphics[width=8.5cm]{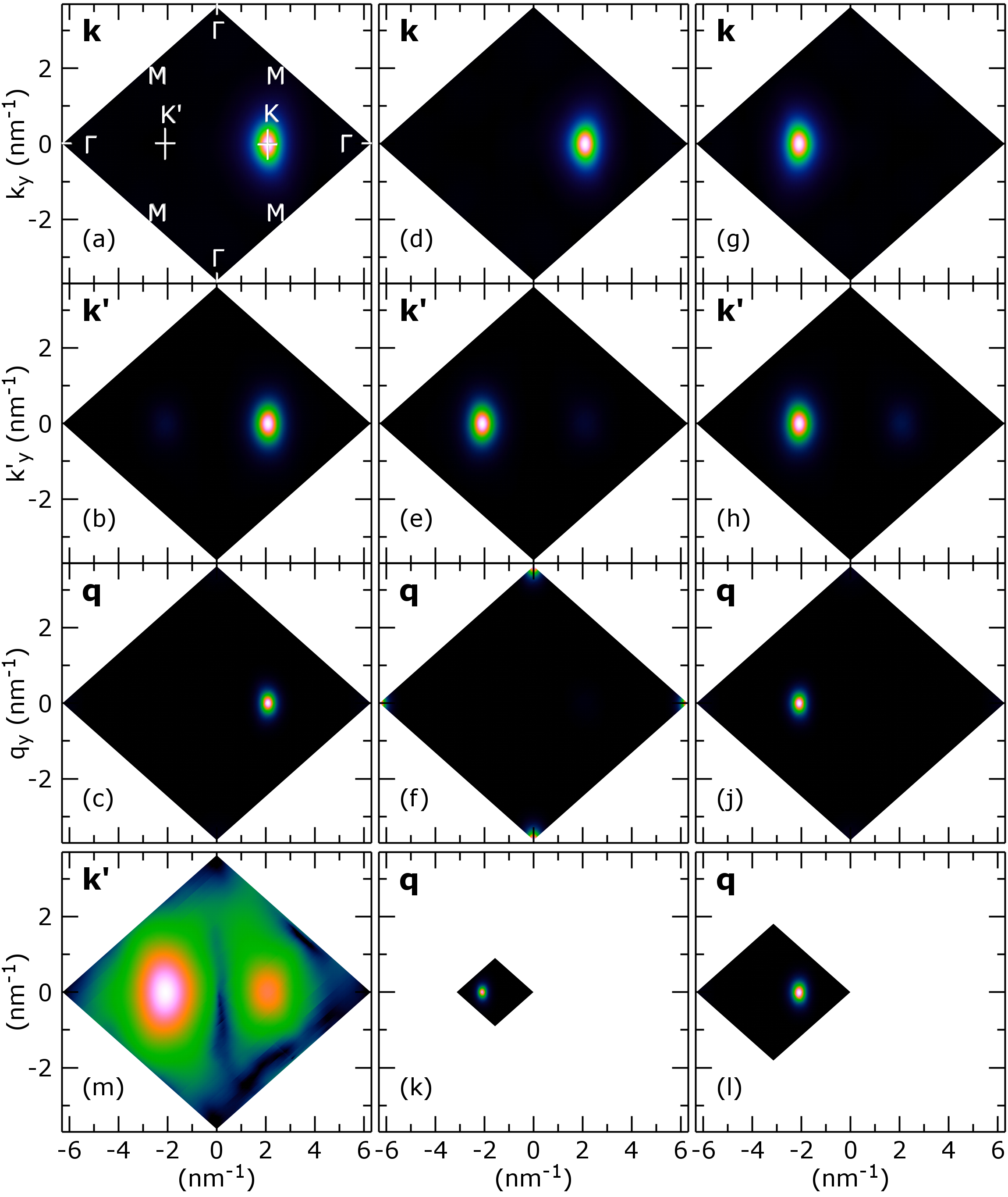}
\caption{Projections of the biexciton wave function to $\mathbf{k}$ (a),(d),(g) $\sum_{\mathbf{k}}\Psi_{h\mathbf{k} e \mathbf{k}+\mathbf{q} h \mathbf{k}'+\mathbf{q} e \mathbf{k}' }$, to $\mathbf{k}'$ (b),(e),(h) $\sum_{\mathbf{k}'}\Psi_{h\mathbf{k} e \mathbf{k}+\mathbf{q} h \mathbf{k}'+\mathbf{q} e\mathbf{k}' }$ or to $\mathbf{q}$ (c),(f),(j)  $\sum_{\mathbf{q}}\Psi_{h\mathbf{k} e \mathbf{k}+\mathbf{q} h \mathbf{k}'+\mathbf{q} e \mathbf{k}' }$.
(a-c) shows the dark biexciton ground state represented by wave function 
$\Psi_{h\downarrow\mathbf{k} e \downarrow\mathbf{k}+\mathbf{q} h \uparrow\mathbf{k}'+\mathbf{q} e \uparrow\mathbf{k}' }$, (d-f) the lowest energy bright biexciton using $\Psi_{h\downarrow\mathbf{k} e \downarrow\mathbf{k}+\mathbf{q} h \uparrow\mathbf{k}'+\mathbf{q} e \uparrow\mathbf{k}' }$  and (g-j) using the wavefunction $\Psi_{h\uparrow\mathbf{k} e \downarrow\mathbf{k}+\mathbf{q} h \downarrow\mathbf{k}'+\mathbf{q} e \uparrow\mathbf{k}' }$ obtained using the strategy from section \ref{convergentexcited}. 
(k) and (l) showing $\mathbf{q}$ restricted to a small and a big box of $\Psi_{h\uparrow\mathbf{k} e \downarrow\mathbf{k}+\mathbf{q} h \downarrow\mathbf{k}'+\mathbf{q} e \uparrow\mathbf{k}' }$. 
(m) shows a logarithmic plot of $\mathbf{k}'$ projection of bright 1s A-B biexciton $\Psi_{h\downarrow\mathbf{k} e \downarrow\mathbf{k}+\mathbf{q} h \downarrow\mathbf{k}'+\mathbf{q} e \downarrow\mathbf{k}' }$ to show the hybridization.}
  \label{biexcitonfig}
\end{figure}

\section{Biexciton states}
\label{biexciton_section}
Biexcitons are electron-holes complexes with one additional particle more than a trion. Biexcitons consists of two electrons and holes and can be viewed as bound states made from two excitons.
The biexciton wavefunction $\Psi_{h\lambda_1\mathbf{k}_1 e \lambda_2\mathbf{k}_2 h \lambda_3\mathbf{k}_3 e \lambda_4\mathbf{k}_4 }$ is calculated using the generalized Bethe-Salpeter Eq. (\ref{bethesalpeterlike}).
 The optical created biexcitons as in the exciton case have overall vanishing momentum, therefore a transformation such as $\Psi_{h\lambda_1\mathbf{k} e \lambda_2\mathbf{k}+\mathbf{q} h \lambda_3\mathbf{k}'+\mathbf{q} e \lambda_4\mathbf{k}' }$ with overall vanishing momentum excludes some dark biexcitons (see Fig. \ref{trionbiexcproj}b) for the tensor network implementation). 
Unfortunately it does not exclude every dark biexciton, since we include the full Brioullin zone (a drawback of the full calculation). For example the typical ground state bright exciton in MoS$_2$ consists of two spin allowed excitons  one at K and one at K' point (cf. Fig. \ref{biexcitonconfig}a) or Fig. \ref{biexcitonfig}(d-f)), however a biexciton made from two spin forbidden excitons one at K and one at K' point (cf. Fig. \ref{biexcitonconfig}b) or Fig. \ref{biexcitonfig}(a-c) ) is included in the same $\Psi_{h\lambda_1\mathbf{k} e \lambda_2\mathbf{k}+\mathbf{q} h \lambda_3\mathbf{k}'+\mathbf{q} e \lambda_4\mathbf{k}' }$ biexciton wavefunction (of course due to small hybridizations between K and K' the dipole moment does not vanish completely but is much smaller). Unfortunately the   dark biexciton, which is uninteresting for optical applications, has a smaller energy than the bright biexciton, so that the retrieval of the bright biexciton for example from ITP (theoretical always retrieving the ground state) is tricky.
One way is to use as starting point for the ITP bright biexciton state calculated with DMRG even with low numerical convergence. 
In principle, the ITP should retrieve the biexciton ground state, but due to limits in the link dimension and variational approach for applying the MPOs often it does not change the overall symmetry of the eigenstate and is stuck inside a local subspace. 
In order to retrieve with DMRG the biexciton state of interest, we often ran the DMRG several times, since for the biexciton case DMRG was often trapped in local minima of the problem, where the symmetry and ordering of the retrieved states depended on the random initial vector.
Furthermore it was also beneficial to use DMRG with  parameters resulting in low convergence, e.g. low link dimension or numerical accuracy, since we obtained  often biexciton states with very different symmetries (e.g. s, p and even d like contributions) in this case.
So the ground state, the dark biexciton is retrieved from a DMRG of $\Psi_{h\downarrow\mathbf{k} e \downarrow\mathbf{k}+\mathbf{q} h \uparrow\mathbf{k}'+\mathbf{q} e \uparrow\mathbf{k}' }$ and subsequent ITP (cf. Fig. \ref{biexcitonfig} (a-c)) yielding an energy of 3.506 eV and thus a binding energy of 27 meV.
Another DMRG run allowed to retrieve the lowest bright biexciton and a subsequent ITP (using $\Psi_{h\downarrow\mathbf{k} e \downarrow\mathbf{k}+\mathbf{q} h \uparrow\mathbf{k}'+\mathbf{q} e \uparrow\mathbf{k}' }$ the strategy from section \ref{convergentexcited}) with energy 3.519 eV and a binding energy of 19-20 meV (cf. Fig. \ref{biexcitonfig}(d-f)).

\SANchange{
Another way to retrieve the bright biexciton is to limit  $\mathbf{q}$ to a box (similar to the  exciton and trion case).
This requires the spin configuration $\Psi_{h\uparrow\mathbf{k} e \downarrow\mathbf{k}+\mathbf{q} h \downarrow\mathbf{k}'+\mathbf{q} e \uparrow\mathbf{k}' }$ together with $\mathbf{q}$ restricted around K' fixing the upper two bits (Fig. \ref{biexcitonfig} k)). 
The calculation yields a
 bright exciton state with energy 3.536 eV and binding energy around 1-2 meV, 
%
which differs considerably from the previous result!
A calculation with a bigger box for $\mathbf{q}$ fixing only one bit results in an energy of 3.523 eV  and binding energy 15 meV (cf. Fig. \ref{biexcitonfig} l)),
%
 but required a higher accuracy for applying UP and link dimension in the seed DMRG. The binding energy is still 5 meV smaller than the previous result, but the major deviation came from a too small box for $\mathbf{q}$ effectively preventing the biexciton binding process (cf. Fig. \ref{biexcitonfig}k) l) ). Furthermore the deviation is roughly twice as large as the error introduced by restricting an exciton to a big box around the K point. 
 Therefore the deviation is caused by a suppression of the K-K' hybridization. This is confirmed by hybridization present in the bright biexciton calculated also for this spin configuration using the method from Sec. \ref{convergentexcited} (cf. Fig.  \ref{biexcitonfig}  (h) ) for the hybridization), which agrees to the previously calculated energy 3.518 eV and binding energy 20 meV and is depicted in Fig. \ref{biexcitonfig} (g-j).

Beside the bright 1s A-A biexciton $\Psi_{h\uparrow\mathbf{k} e \downarrow\mathbf{k}+\mathbf{q} h \downarrow\mathbf{k}'+\mathbf{q} e \uparrow\mathbf{k}' }$ also allows to retrieve the bright 1s B-B biexciton (cf. Fig. \ref{biexcitonconfig}d)) with 3.791 eV  and 21 meV binding energy.

Another interesting state is the bright 1s A-B biexciton, which can be retrieved from $\Psi_{h\downarrow\mathbf{k} e \downarrow\mathbf{k}+\mathbf{q} h \downarrow\mathbf{k}'+\mathbf{q} e \downarrow\mathbf{k}' }$.
Two states are retrieved one with $\mathbf{k}$, $\mathbf{k}'$, $\mathbf{q}$ around K and one around K', both with the energy 3.655 eV and 20 meV binding energy  within numerical accuracy.
These are actually not two different eigenstates, but the same eigenstate, since the ansatz using only the expansion coefficients $\Psi_{h\downarrow\mathbf{k} e \downarrow\mathbf{k}+\mathbf{q} h \downarrow\mathbf{k}'+\mathbf{q} e \downarrow\mathbf{k}' }$ does not guarantee the correct permutation symmetry. 
Therefore a linear combination of the two degenerate results yields biexciton state with the proper symmetrization.
This does not effect the eigenenergies, but for a calculation of matrix elements, proper symmetrization is required using the expansion coefficients from the calculation presented here.

Furthermore for all biexciton states a small hybridization for the hole between K and K' is visible for $\mathbf{k}'$ (cf. Fig. \ref{biexcitonfig} (b), (e), (h) and (m)), if $\mathbf{q}$ is not restricted.
We yield in all cases almost the same energy, since our model system is lacking short range exchange interaction \cite{steinhoff2018biexciton}.
 
}

 \begin{figure}[tb] 	
 \centering
   \includegraphics[width=8.0cm]{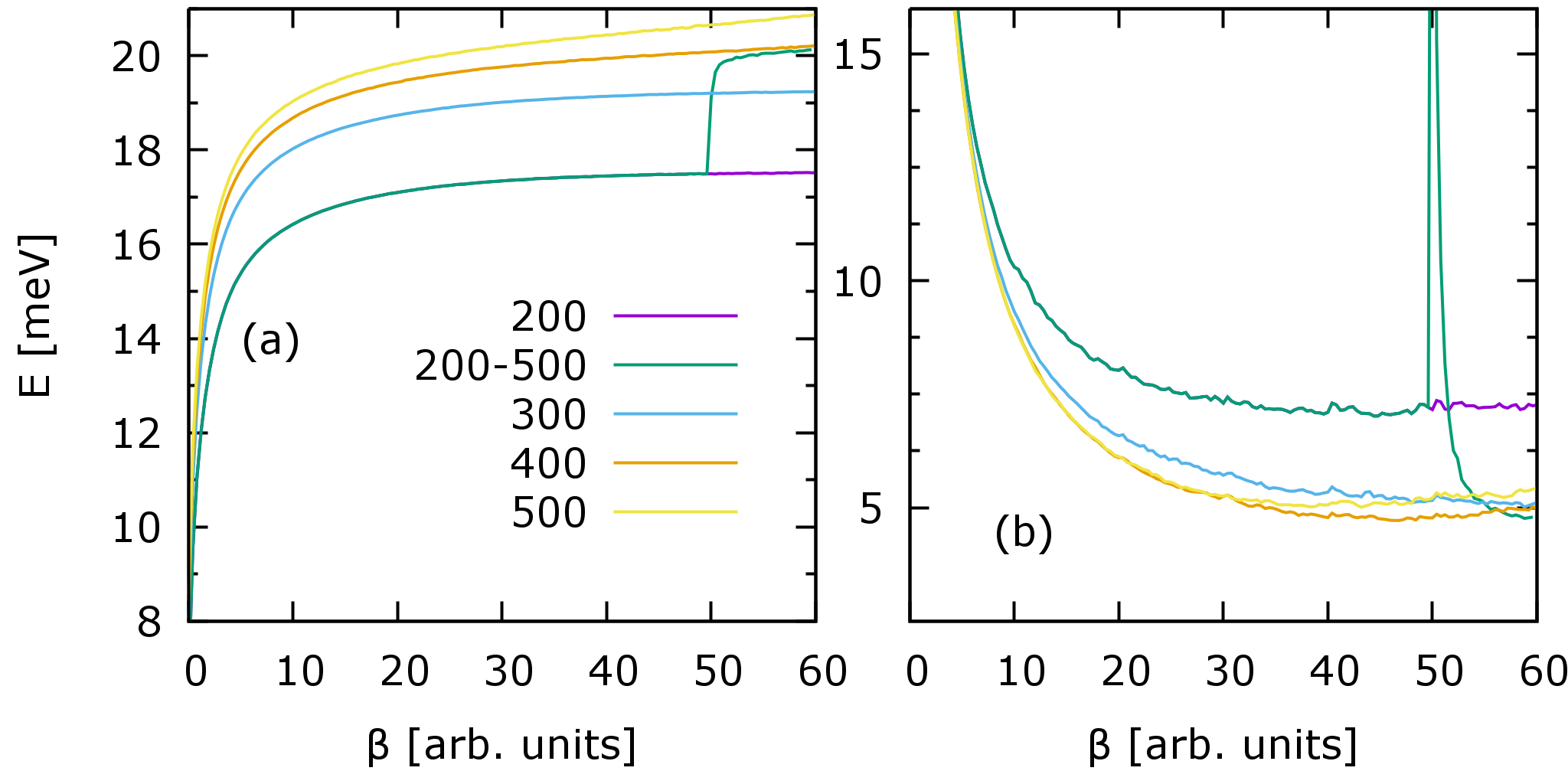}
   \caption{ ITP of the bright biexciton state in MoS$_2$ a) $\langle E \rangle$ binding energy  and b) $\langle \Delta E \rangle$ energy variation   over imaginary time $\beta$. }
   \label{figbiexcprop}
 \end{figure}
\subsection{Convergent and fast calculations}
\label{convergentexcited}
The link dimension of the MPS appearing during ITP for trion and biexcitons for a truncation precision of $10^{-12}$ exceeds the feasible range.
Therefore the link dimension is truncated, which due to the polynomial scaling of the computation time and memory leads to a significant speed up.
The energy $\langle E\rangle$ and energy variation $\langle \Delta E \rangle$ is plotted over propagation time $\beta$ in Fig. \ref{figbiexcprop} a) and b) for the bright biexciton state (note the dark biexciton state is the ground state). 
The energy in Fig. \ref{figbiexcprop} a) converges at least for the low link dimension (up to 300) for increasing, also for $\beta<30$  $\langle E \rangle$ clearly converges for increasing link dimension.
But for $\beta>30$ the $\langle E\rangle$ for 300 and 400 link dimension does not seem to converge for $\beta\rightarrow\infty$ and also increasing link dimension does not seem to lead to convergence.
Decreasing $\langle \Delta E \rangle$ in Fig. \ref{figbiexcprop} is an indicator for convergence and for up to link dimension 400 it goes to a fixed value (for biexcitons typically 5 meV, for excitons and trions typically 0.5-2 meV ).  But $\langle \Delta E \rangle$ is actually slightly increasing for link dimension 400 and 500 a clear sign that the propagation is leaving the initial bright biexciton state and propagating towards the (dark) biexciton ground state. For lower link dimensions the subspace of the propagation is so small, that the MPS sticks to the type of the  initially prepared state. 
This explains the convergence problems visible in Fig. \ref{figbiexcprop} after $\beta=30$ and of course does not occur, if we calculate the ground state within the projector $UP$.

In order to keep the initial state, which is not the ground state, it is better to raise the link dimension relatively late during ITP for a relative short remaining propagation time $\beta$ (see Fig. \ref{figbiexcprop} link 200-500). In this way the calculation benefits from the fast calculation with small link dimension and the small propagation subspace in the beginning and the good energy convergence for the high link dimension at the very end. Therefore this strategy was used for most trion and biexciton calculations in this paper. 

Please note that $\langle \Delta E \rangle$ is not necessarily an absolute  measure of the error $\langle E \rangle$ of the eigenstate energy, but about the quality of the wave function (admixture from other eigenstates) and that it is also affected by a truncated link dimension.

\section{Conclusion}
Detailed strategies, algorithm and background information for an efficient calculation of bound electron hole complexes in two dimensional material such as MoS$_2$ using tensor network methods were given in this paper.
Furthermore example calculations for exciton, trions, biexcitons showed how to selectively retrieve specific bound states.
We believe, that the tensor methods with logical circuits will  enable the  calculating bound electron hole complexes on the full Brioullin zone with high resolution grid discretization for many future applications.

\begin{acknowledgments}
\noindent We gratefully acknowledge support from the Deutsche Forschungsgemeinschaft (DFG) through SFB 787 B1 (Project No.  43659573).
\end{acknowledgments}



\begin{thebibliography}{65}%
\makeatletter
\providecommand \@ifxundefined [1]{%
 \@ifx{#1\undefined}
}%
\providecommand \@ifnum [1]{%
 \ifnum #1\expandafter \@firstoftwo
 \else \expandafter \@secondoftwo
 \fi
}%
\providecommand \@ifx [1]{%
 \ifx #1\expandafter \@firstoftwo
 \else \expandafter \@secondoftwo
 \fi
}%
\providecommand \natexlab [1]{#1}%
\providecommand \enquote  [1]{``#1''}%
\providecommand \bibnamefont  [1]{#1}%
\providecommand \bibfnamefont [1]{#1}%
\providecommand \citenamefont [1]{#1}%
\providecommand \href@noop [0]{\@secondoftwo}%
\providecommand \href [0]{\begingroup \@sanitize@url \@href}%
\providecommand \@href[1]{\@@startlink{#1}\@@href}%
\providecommand \@@href[1]{\endgroup#1\@@endlink}%
\providecommand \@sanitize@url [0]{\catcode `\\12\catcode `\$12\catcode
  `\&12\catcode `\#12\catcode `\^12\catcode `\_12\catcode `\%12\relax}%
\providecommand \@@startlink[1]{}%
\providecommand \@@endlink[0]{}%
\providecommand \url  [0]{\begingroup\@sanitize@url \@url }%
\providecommand \@url [1]{\endgroup\@href {#1}{\urlprefix }}%
\providecommand \urlprefix  [0]{URL }%
\providecommand \Eprint [0]{\href }%
\providecommand \doibase [0]{http://dx.doi.org/}%
\providecommand \selectlanguage [0]{\@gobble}%
\providecommand \bibinfo  [0]{\@secondoftwo}%
\providecommand \bibfield  [0]{\@secondoftwo}%
\providecommand \translation [1]{[#1]}%
\providecommand \BibitemOpen [0]{}%
\providecommand \bibitemStop [0]{}%
\providecommand \bibitemNoStop [0]{.\EOS\space}%
\providecommand \EOS [0]{\spacefactor3000\relax}%
\providecommand \BibitemShut  [1]{\csname bibitem#1\endcsname}%
\let\auto@bib@innerbib\@empty
\bibitem [{\citenamefont {Lindberg}\ and\ \citenamefont
  {Koch}(1988)}]{lindberg1988effective}%
  \BibitemOpen
  \bibfield  {author} {\bibinfo {author} {\bibfnamefont {M.}~\bibnamefont
  {Lindberg}}\ and\ \bibinfo {author} {\bibfnamefont {S.~W.}\ \bibnamefont
  {Koch}},\ }\href@noop {} {\bibfield  {journal} {\bibinfo  {journal} {Physical
  Review B}\ }\textbf {\bibinfo {volume} {38}},\ \bibinfo {pages} {3342}
  (\bibinfo {year} {1988})}\BibitemShut {NoStop}%
\bibitem [{\citenamefont {Butscher}\ \emph {et~al.}(2005)\citenamefont
  {Butscher}, \citenamefont {F{\"o}rstner}, \citenamefont {Waldm{\"u}ller},\
  and\ \citenamefont {Knorr}}]{butscher2005ultrafast}%
  \BibitemOpen
  \bibfield  {author} {\bibinfo {author} {\bibfnamefont {S.}~\bibnamefont
  {Butscher}}, \bibinfo {author} {\bibfnamefont {J.}~\bibnamefont
  {F{\"o}rstner}}, \bibinfo {author} {\bibfnamefont {I.}~\bibnamefont
  {Waldm{\"u}ller}}, \ and\ \bibinfo {author} {\bibfnamefont {A.}~\bibnamefont
  {Knorr}},\ }\href@noop {} {\bibfield  {journal} {\bibinfo  {journal}
  {Physical Review B}\ }\textbf {\bibinfo {volume} {72}},\ \bibinfo {pages}
  {045314} (\bibinfo {year} {2005})}\BibitemShut {NoStop}%
\bibitem [{\citenamefont {Malić}\ \emph {et~al.}(2008)\citenamefont {Malić},
  \citenamefont {Hirtschulz}, \citenamefont {Milde}, \citenamefont {Richter},
  \citenamefont {Maultzsch}, \citenamefont {Reich},\ and\ \citenamefont
  {Knorr}}]{doi:10.1002/pssb.200879612}%
  \BibitemOpen
  \bibfield  {author} {\bibinfo {author} {\bibfnamefont {E.}~\bibnamefont
  {Malić}}, \bibinfo {author} {\bibfnamefont {M.}~\bibnamefont {Hirtschulz}},
  \bibinfo {author} {\bibfnamefont {F.}~\bibnamefont {Milde}}, \bibinfo
  {author} {\bibfnamefont {M.}~\bibnamefont {Richter}}, \bibinfo {author}
  {\bibfnamefont {J.}~\bibnamefont {Maultzsch}}, \bibinfo {author}
  {\bibfnamefont {S.}~\bibnamefont {Reich}}, \ and\ \bibinfo {author}
  {\bibfnamefont {A.}~\bibnamefont {Knorr}},\ }\href {\doibase
  10.1002/pssb.200879612} {\bibfield  {journal} {\bibinfo  {journal} {physica
  status solidi (b)}\ }\textbf {\bibinfo {volume} {245}},\ \bibinfo {pages}
  {2155} (\bibinfo {year} {2008})}\BibitemShut {NoStop}%
\bibitem [{\citenamefont {Reiter}\ \emph {et~al.}(2007)\citenamefont {Reiter},
  \citenamefont {Glanemann}, \citenamefont {Axt},\ and\ \citenamefont
  {Kuhn}}]{reiter2007spatiotemporal}%
  \BibitemOpen
  \bibfield  {author} {\bibinfo {author} {\bibfnamefont {D.}~\bibnamefont
  {Reiter}}, \bibinfo {author} {\bibfnamefont {M.}~\bibnamefont {Glanemann}},
  \bibinfo {author} {\bibfnamefont {V.}~\bibnamefont {Axt}}, \ and\ \bibinfo
  {author} {\bibfnamefont {T.}~\bibnamefont {Kuhn}},\ }\href@noop {} {\bibfield
   {journal} {\bibinfo  {journal} {Physical Review B}\ }\textbf {\bibinfo
  {volume} {75}},\ \bibinfo {pages} {205327} (\bibinfo {year}
  {2007})}\BibitemShut {NoStop}%
\bibitem [{\citenamefont {Schmidt}\ \emph {et~al.}(2016)\citenamefont
  {Schmidt}, \citenamefont {Bergh{\"a}user}, \citenamefont {Schneider},
  \citenamefont {Selig}, \citenamefont {Tonndorf}, \citenamefont {Malic},
  \citenamefont {Knorr}, \citenamefont {Michaelis~de Vasconcellos},\ and\
  \citenamefont {Bratschitsch}}]{schmidt2016ultrafast}%
  \BibitemOpen
  \bibfield  {author} {\bibinfo {author} {\bibfnamefont {R.}~\bibnamefont
  {Schmidt}}, \bibinfo {author} {\bibfnamefont {G.}~\bibnamefont
  {Bergh{\"a}user}}, \bibinfo {author} {\bibfnamefont {R.}~\bibnamefont
  {Schneider}}, \bibinfo {author} {\bibfnamefont {M.}~\bibnamefont {Selig}},
  \bibinfo {author} {\bibfnamefont {P.}~\bibnamefont {Tonndorf}}, \bibinfo
  {author} {\bibfnamefont {E.}~\bibnamefont {Malic}}, \bibinfo {author}
  {\bibfnamefont {A.}~\bibnamefont {Knorr}}, \bibinfo {author} {\bibfnamefont
  {S.}~\bibnamefont {Michaelis~de Vasconcellos}}, \ and\ \bibinfo {author}
  {\bibfnamefont {R.}~\bibnamefont {Bratschitsch}},\ }\href {\doibase
  10.1021/acs.nanolett.5b04733} {\bibfield  {journal} {\bibinfo  {journal}
  {Nano Letters}\ }\textbf {\bibinfo {volume} {16}},\ \bibinfo {pages} {2945}
  (\bibinfo {year} {2016})},\ \bibinfo {note} {pMID: 27086935},\ \Eprint
  {http://arxiv.org/abs/https://doi.org/10.1021/acs.nanolett.5b04733}
  {https://doi.org/10.1021/acs.nanolett.5b04733} \BibitemShut {NoStop}%
\bibitem [{\citenamefont {Chatterjee}\ \emph {et~al.}(2004)\citenamefont
  {Chatterjee}, \citenamefont {Ell}, \citenamefont {Mosor}, \citenamefont
  {Khitrova}, \citenamefont {Gibbs}, \citenamefont {Hoyer}, \citenamefont
  {Kira}, \citenamefont {Koch}, \citenamefont {Prineas},\ and\ \citenamefont
  {Stolz}}]{PhysRevLett.92.067402}%
  \BibitemOpen
  \bibfield  {author} {\bibinfo {author} {\bibfnamefont {S.}~\bibnamefont
  {Chatterjee}}, \bibinfo {author} {\bibfnamefont {C.}~\bibnamefont {Ell}},
  \bibinfo {author} {\bibfnamefont {S.}~\bibnamefont {Mosor}}, \bibinfo
  {author} {\bibfnamefont {G.}~\bibnamefont {Khitrova}}, \bibinfo {author}
  {\bibfnamefont {H.~M.}\ \bibnamefont {Gibbs}}, \bibinfo {author}
  {\bibfnamefont {W.}~\bibnamefont {Hoyer}}, \bibinfo {author} {\bibfnamefont
  {M.}~\bibnamefont {Kira}}, \bibinfo {author} {\bibfnamefont {S.~W.}\
  \bibnamefont {Koch}}, \bibinfo {author} {\bibfnamefont {J.~P.}\ \bibnamefont
  {Prineas}}, \ and\ \bibinfo {author} {\bibfnamefont {H.}~\bibnamefont
  {Stolz}},\ }\href {\doibase 10.1103/PhysRevLett.92.067402} {\bibfield
  {journal} {\bibinfo  {journal} {Phys. Rev. Lett.}\ }\textbf {\bibinfo
  {volume} {92}},\ \bibinfo {pages} {067402} (\bibinfo {year}
  {2004})}\BibitemShut {NoStop}%
\bibitem [{\citenamefont {Winzer}\ \emph {et~al.}(2010)\citenamefont {Winzer},
  \citenamefont {Knorr},\ and\ \citenamefont {Malic}}]{winzer2010carrier}%
  \BibitemOpen
  \bibfield  {author} {\bibinfo {author} {\bibfnamefont {T.}~\bibnamefont
  {Winzer}}, \bibinfo {author} {\bibfnamefont {A.}~\bibnamefont {Knorr}}, \
  and\ \bibinfo {author} {\bibfnamefont {E.}~\bibnamefont {Malic}},\
  }\href@noop {} {\bibfield  {journal} {\bibinfo  {journal} {Nano letters}\
  }\textbf {\bibinfo {volume} {10}},\ \bibinfo {pages} {4839} (\bibinfo {year}
  {2010})}\BibitemShut {NoStop}%
\bibitem [{\citenamefont {Meckbach}\ \emph {et~al.}(2018)\citenamefont
  {Meckbach}, \citenamefont {Stroucken},\ and\ \citenamefont
  {Koch}}]{PhysRevB.97.035425}%
  \BibitemOpen
  \bibfield  {author} {\bibinfo {author} {\bibfnamefont {L.}~\bibnamefont
  {Meckbach}}, \bibinfo {author} {\bibfnamefont {T.}~\bibnamefont {Stroucken}},
  \ and\ \bibinfo {author} {\bibfnamefont {S.~W.}\ \bibnamefont {Koch}},\
  }\href {\doibase 10.1103/PhysRevB.97.035425} {\bibfield  {journal} {\bibinfo
  {journal} {Phys. Rev. B}\ }\textbf {\bibinfo {volume} {97}},\ \bibinfo
  {pages} {035425} (\bibinfo {year} {2018})}\BibitemShut {NoStop}%
\bibitem [{\citenamefont {Qiu}\ \emph {et~al.}(2013{\natexlab{a}})\citenamefont
  {Qiu}, \citenamefont {Felipe},\ and\ \citenamefont {Louie}}]{qiu2013optical}%
  \BibitemOpen
  \bibfield  {author} {\bibinfo {author} {\bibfnamefont {D.~Y.}\ \bibnamefont
  {Qiu}}, \bibinfo {author} {\bibfnamefont {H.}~\bibnamefont {Felipe}}, \ and\
  \bibinfo {author} {\bibfnamefont {S.~G.}\ \bibnamefont {Louie}},\ }\href@noop
  {} {\bibfield  {journal} {\bibinfo  {journal} {Phys. Rev. Lett.}\ }\textbf
  {\bibinfo {volume} {111}},\ \bibinfo {pages} {216805} (\bibinfo {year}
  {2013}{\natexlab{a}})}\BibitemShut {NoStop}%
\bibitem [{\citenamefont {Bergh{\"a}user}\ and\ \citenamefont
  {Malic}(2014)}]{berghauser2014analytical}%
  \BibitemOpen
  \bibfield  {author} {\bibinfo {author} {\bibfnamefont {G.}~\bibnamefont
  {Bergh{\"a}user}}\ and\ \bibinfo {author} {\bibfnamefont {E.}~\bibnamefont
  {Malic}},\ }\href@noop {} {\bibfield  {journal} {\bibinfo  {journal} {Phys.
  Rev. B}\ }\textbf {\bibinfo {volume} {89}},\ \bibinfo {pages} {125309}
  (\bibinfo {year} {2014})}\BibitemShut {NoStop}%
\bibitem [{\citenamefont {Qiu}\ \emph {et~al.}(2015)\citenamefont {Qiu},
  \citenamefont {Cao},\ and\ \citenamefont {Louie}}]{qiu2015nonanalyticity}%
  \BibitemOpen
  \bibfield  {author} {\bibinfo {author} {\bibfnamefont {D.~Y.}\ \bibnamefont
  {Qiu}}, \bibinfo {author} {\bibfnamefont {T.}~\bibnamefont {Cao}}, \ and\
  \bibinfo {author} {\bibfnamefont {S.~G.}\ \bibnamefont {Louie}},\ }\href@noop
  {} {\bibfield  {journal} {\bibinfo  {journal} {Physical review letters}\
  }\textbf {\bibinfo {volume} {115}},\ \bibinfo {pages} {176801} (\bibinfo
  {year} {2015})}\BibitemShut {NoStop}%
\bibitem [{\citenamefont {Haug}\ and\ \citenamefont
  {Koch}(2009)}]{haug2009quantum}%
  \BibitemOpen
  \bibfield  {author} {\bibinfo {author} {\bibfnamefont {H.}~\bibnamefont
  {Haug}}\ and\ \bibinfo {author} {\bibfnamefont {S.~W.}\ \bibnamefont
  {Koch}},\ }\href@noop {} {\emph {\bibinfo {title} {Quantum Theory of the
  Optical and Electronic Properties of Semiconductors: Fivth Edition}}}\
  (\bibinfo  {publisher} {World Scientific Publishing Company},\ \bibinfo
  {year} {2009})\BibitemShut {NoStop}%
\bibitem [{\citenamefont {H{\"u}ser}\ \emph {et~al.}(2013)\citenamefont
  {H{\"u}ser}, \citenamefont {Olsen},\ and\ \citenamefont
  {Thygesen}}]{huser2013dielectric}%
  \BibitemOpen
  \bibfield  {author} {\bibinfo {author} {\bibfnamefont {F.}~\bibnamefont
  {H{\"u}ser}}, \bibinfo {author} {\bibfnamefont {T.}~\bibnamefont {Olsen}}, \
  and\ \bibinfo {author} {\bibfnamefont {K.~S.}\ \bibnamefont {Thygesen}},\
  }\href@noop {} {\bibfield  {journal} {\bibinfo  {journal} {Physical Review
  B}\ }\textbf {\bibinfo {volume} {88}},\ \bibinfo {pages} {245309} (\bibinfo
  {year} {2013})}\BibitemShut {NoStop}%
\bibitem [{\citenamefont {Ridolfi}\ \emph {et~al.}(2018)\citenamefont
  {Ridolfi}, \citenamefont {Lewenkopf},\ and\ \citenamefont
  {Pereira}}]{ridolfi2018exstruc}%
  \BibitemOpen
  \bibfield  {author} {\bibinfo {author} {\bibfnamefont {E.}~\bibnamefont
  {Ridolfi}}, \bibinfo {author} {\bibfnamefont {C.~H.}\ \bibnamefont
  {Lewenkopf}}, \ and\ \bibinfo {author} {\bibfnamefont {V.~M.}\ \bibnamefont
  {Pereira}},\ }\href {\doibase 10.1103/PhysRevB.97.205409} {\bibfield
  {journal} {\bibinfo  {journal} {Phys. Rev. B}\ }\textbf {\bibinfo {volume}
  {97}},\ \bibinfo {pages} {205409} (\bibinfo {year} {2018})}\BibitemShut
  {NoStop}%
\bibitem [{\citenamefont {Deilmann}\ and\ \citenamefont
  {Thygesen}(2019)}]{deilmann2019finite}%
  \BibitemOpen
  \bibfield  {author} {\bibinfo {author} {\bibfnamefont {T.}~\bibnamefont
  {Deilmann}}\ and\ \bibinfo {author} {\bibfnamefont {K.~S.}\ \bibnamefont
  {Thygesen}},\ }\href@noop {} {\bibfield  {journal} {\bibinfo  {journal} {2D
  Materials}\ }\textbf {\bibinfo {volume} {6}},\ \bibinfo {pages} {035003}
  (\bibinfo {year} {2019})}\BibitemShut {NoStop}%
\bibitem [{\citenamefont {Steinhoff}\ \emph
  {et~al.}(2018{\natexlab{a}})\citenamefont {Steinhoff}, \citenamefont
  {Wehling},\ and\ \citenamefont {R{\"o}sner}}]{steinhoff2018frequency}%
  \BibitemOpen
  \bibfield  {author} {\bibinfo {author} {\bibfnamefont {A.}~\bibnamefont
  {Steinhoff}}, \bibinfo {author} {\bibfnamefont {T.}~\bibnamefont {Wehling}},
  \ and\ \bibinfo {author} {\bibfnamefont {M.}~\bibnamefont {R{\"o}sner}},\
  }\href@noop {} {\bibfield  {journal} {\bibinfo  {journal} {Physical Review
  B}\ }\textbf {\bibinfo {volume} {98}},\ \bibinfo {pages} {045304} (\bibinfo
  {year} {2018}{\natexlab{a}})}\BibitemShut {NoStop}%
\bibitem [{\citenamefont {St{\'e}b{\'e}}\ \emph {et~al.}(1997)\citenamefont
  {St{\'e}b{\'e}}, \citenamefont {Munschy}, \citenamefont {Stauffer},
  \citenamefont {Dujardin},\ and\ \citenamefont {Murat}}]{stebe1997excitonic}%
  \BibitemOpen
  \bibfield  {author} {\bibinfo {author} {\bibfnamefont {B.}~\bibnamefont
  {St{\'e}b{\'e}}}, \bibinfo {author} {\bibfnamefont {G.}~\bibnamefont
  {Munschy}}, \bibinfo {author} {\bibfnamefont {L.}~\bibnamefont {Stauffer}},
  \bibinfo {author} {\bibfnamefont {F.}~\bibnamefont {Dujardin}}, \ and\
  \bibinfo {author} {\bibfnamefont {J.}~\bibnamefont {Murat}},\ }\href@noop {}
  {\bibfield  {journal} {\bibinfo  {journal} {Physical Review B}\ }\textbf
  {\bibinfo {volume} {56}},\ \bibinfo {pages} {12454} (\bibinfo {year}
  {1997})}\BibitemShut {NoStop}%
\bibitem [{\citenamefont {Mayrock}\ \emph {et~al.}(1999)\citenamefont
  {Mayrock}, \citenamefont {W{\"u}nsche}, \citenamefont {Henneberger},
  \citenamefont {Riva}, \citenamefont {Schweigert},\ and\ \citenamefont
  {Peeters}}]{mayrock1999weak}%
  \BibitemOpen
  \bibfield  {author} {\bibinfo {author} {\bibfnamefont {O.}~\bibnamefont
  {Mayrock}}, \bibinfo {author} {\bibfnamefont {H.-J.}\ \bibnamefont
  {W{\"u}nsche}}, \bibinfo {author} {\bibfnamefont {F.}~\bibnamefont
  {Henneberger}}, \bibinfo {author} {\bibfnamefont {C.}~\bibnamefont {Riva}},
  \bibinfo {author} {\bibfnamefont {V.}~\bibnamefont {Schweigert}}, \ and\
  \bibinfo {author} {\bibfnamefont {F.}~\bibnamefont {Peeters}},\ }\href@noop
  {} {\bibfield  {journal} {\bibinfo  {journal} {Physical Review B}\ }\textbf
  {\bibinfo {volume} {60}},\ \bibinfo {pages} {5582} (\bibinfo {year}
  {1999})}\BibitemShut {NoStop}%
\bibitem [{\citenamefont {Esser}\ \emph {et~al.}(2000)\citenamefont {Esser},
  \citenamefont {Runge}, \citenamefont {Zimmermann},\ and\ \citenamefont
  {Langbein}}]{zimmermanntrion2000}%
  \BibitemOpen
  \bibfield  {author} {\bibinfo {author} {\bibfnamefont {A.}~\bibnamefont
  {Esser}}, \bibinfo {author} {\bibfnamefont {E.}~\bibnamefont {Runge}},
  \bibinfo {author} {\bibfnamefont {R.}~\bibnamefont {Zimmermann}}, \ and\
  \bibinfo {author} {\bibfnamefont {W.}~\bibnamefont {Langbein}},\ }\href
  {\doibase 10.1002/1521-396X(200003)178:1<489::AID-PSSA489>3.0.CO;2-R}
  {\bibfield  {journal} {\bibinfo  {journal} {physica status solidi (a)}\
  }\textbf {\bibinfo {volume} {178}},\ \bibinfo {pages} {489} (\bibinfo {year}
  {2000})}\BibitemShut {NoStop}%
\bibitem [{\citenamefont {Esser}\ \emph {et~al.}(2001)\citenamefont {Esser},
  \citenamefont {Zimmermann},\ and\ \citenamefont {Runge}}]{esser2001theory}%
  \BibitemOpen
  \bibfield  {author} {\bibinfo {author} {\bibfnamefont {A.}~\bibnamefont
  {Esser}}, \bibinfo {author} {\bibfnamefont {R.}~\bibnamefont {Zimmermann}}, \
  and\ \bibinfo {author} {\bibfnamefont {E.}~\bibnamefont {Runge}},\
  }\href@noop {} {\bibfield  {journal} {\bibinfo  {journal} {physica status
  solidi (b)}\ }\textbf {\bibinfo {volume} {227}},\ \bibinfo {pages} {317}
  (\bibinfo {year} {2001})}\BibitemShut {NoStop}%
\bibitem [{\citenamefont {Mostaani}\ \emph {et~al.}(2017)\citenamefont
  {Mostaani}, \citenamefont {Szyniszewski}, \citenamefont {Price},
  \citenamefont {Maezono}, \citenamefont {Danovich}, \citenamefont {Hunt},
  \citenamefont {Drummond},\ and\ \citenamefont
  {Fal'Ko}}]{mostaani2017diffusion}%
  \BibitemOpen
  \bibfield  {author} {\bibinfo {author} {\bibfnamefont {E.}~\bibnamefont
  {Mostaani}}, \bibinfo {author} {\bibfnamefont {M.}~\bibnamefont
  {Szyniszewski}}, \bibinfo {author} {\bibfnamefont {C.}~\bibnamefont {Price}},
  \bibinfo {author} {\bibfnamefont {R.}~\bibnamefont {Maezono}}, \bibinfo
  {author} {\bibfnamefont {M.}~\bibnamefont {Danovich}}, \bibinfo {author}
  {\bibfnamefont {R.}~\bibnamefont {Hunt}}, \bibinfo {author} {\bibfnamefont
  {N.}~\bibnamefont {Drummond}}, \ and\ \bibinfo {author} {\bibfnamefont
  {V.}~\bibnamefont {Fal'Ko}},\ }\href@noop {} {\bibfield  {journal} {\bibinfo
  {journal} {Physical Review B}\ }\textbf {\bibinfo {volume} {96}},\ \bibinfo
  {pages} {075431} (\bibinfo {year} {2017})}\BibitemShut {NoStop}%
\bibitem [{\citenamefont {Singh}\ \emph {et~al.}(2016)\citenamefont {Singh},
  \citenamefont {Moody}, \citenamefont {Tran}, \citenamefont {Scott},
  \citenamefont {Overbeck}, \citenamefont {Bergh\"auser}, \citenamefont
  {Schaibley}, \citenamefont {Seifert}, \citenamefont {Pleskot}, \citenamefont
  {Gabor}, \citenamefont {Yan}, \citenamefont {Mandrus}, \citenamefont
  {Richter}, \citenamefont {Malic}, \citenamefont {Xu},\ and\ \citenamefont
  {Li}}]{PhysRevB.93.041401}%
  \BibitemOpen
  \bibfield  {author} {\bibinfo {author} {\bibfnamefont {A.}~\bibnamefont
  {Singh}}, \bibinfo {author} {\bibfnamefont {G.}~\bibnamefont {Moody}},
  \bibinfo {author} {\bibfnamefont {K.}~\bibnamefont {Tran}}, \bibinfo {author}
  {\bibfnamefont {M.~E.}\ \bibnamefont {Scott}}, \bibinfo {author}
  {\bibfnamefont {V.}~\bibnamefont {Overbeck}}, \bibinfo {author}
  {\bibfnamefont {G.}~\bibnamefont {Bergh\"auser}}, \bibinfo {author}
  {\bibfnamefont {J.}~\bibnamefont {Schaibley}}, \bibinfo {author}
  {\bibfnamefont {E.~J.}\ \bibnamefont {Seifert}}, \bibinfo {author}
  {\bibfnamefont {D.}~\bibnamefont {Pleskot}}, \bibinfo {author} {\bibfnamefont
  {N.~M.}\ \bibnamefont {Gabor}}, \bibinfo {author} {\bibfnamefont
  {J.}~\bibnamefont {Yan}}, \bibinfo {author} {\bibfnamefont {D.~G.}\
  \bibnamefont {Mandrus}}, \bibinfo {author} {\bibfnamefont {M.}~\bibnamefont
  {Richter}}, \bibinfo {author} {\bibfnamefont {E.}~\bibnamefont {Malic}},
  \bibinfo {author} {\bibfnamefont {X.}~\bibnamefont {Xu}}, \ and\ \bibinfo
  {author} {\bibfnamefont {X.}~\bibnamefont {Li}},\ }\href {\doibase
  10.1103/PhysRevB.93.041401} {\bibfield  {journal} {\bibinfo  {journal} {Phys.
  Rev. B}\ }\textbf {\bibinfo {volume} {93}},\ \bibinfo {pages} {041401}
  (\bibinfo {year} {2016})}\BibitemShut {NoStop}%
\bibitem [{\citenamefont {Florian}\ \emph {et~al.}(2018)\citenamefont
  {Florian}, \citenamefont {Hartmann}, \citenamefont {Steinhoff}, \citenamefont
  {Klein}, \citenamefont {Holleitner}, \citenamefont {Finley}, \citenamefont
  {Wehling}, \citenamefont {Kaniber},\ and\ \citenamefont
  {Gies}}]{10.1021acs.nanolett.8b00840}%
  \BibitemOpen
  \bibfield  {author} {\bibinfo {author} {\bibfnamefont {M.}~\bibnamefont
  {Florian}}, \bibinfo {author} {\bibfnamefont {M.}~\bibnamefont {Hartmann}},
  \bibinfo {author} {\bibfnamefont {A.}~\bibnamefont {Steinhoff}}, \bibinfo
  {author} {\bibfnamefont {J.}~\bibnamefont {Klein}}, \bibinfo {author}
  {\bibfnamefont {A.~W.}\ \bibnamefont {Holleitner}}, \bibinfo {author}
  {\bibfnamefont {J.~J.}\ \bibnamefont {Finley}}, \bibinfo {author}
  {\bibfnamefont {T.~O.}\ \bibnamefont {Wehling}}, \bibinfo {author}
  {\bibfnamefont {M.}~\bibnamefont {Kaniber}}, \ and\ \bibinfo {author}
  {\bibfnamefont {C.}~\bibnamefont {Gies}},\ }\href {\doibase
  10.1021/acs.nanolett.8b00840} {\bibfield  {journal} {\bibinfo  {journal}
  {Nano Letters}\ }\textbf {\bibinfo {volume} {18}},\ \bibinfo {pages} {2725}
  (\bibinfo {year} {2018})},\ \bibinfo {note} {pMID: 29558797},\ \Eprint
  {http://arxiv.org/abs/https://doi.org/10.1021/acs.nanolett.8b00840}
  {https://doi.org/10.1021/acs.nanolett.8b00840} \BibitemShut {NoStop}%
\bibitem [{\citenamefont {Zhang}\ \emph {et~al.}(2015)\citenamefont {Zhang},
  \citenamefont {Kidd},\ and\ \citenamefont
  {Varga}}]{doi:10.1021/acs.nanolett.5b03009}%
  \BibitemOpen
  \bibfield  {author} {\bibinfo {author} {\bibfnamefont {D.~K.}\ \bibnamefont
  {Zhang}}, \bibinfo {author} {\bibfnamefont {D.~W.}\ \bibnamefont {Kidd}}, \
  and\ \bibinfo {author} {\bibfnamefont {K.}~\bibnamefont {Varga}},\ }\href
  {\doibase 10.1021/acs.nanolett.5b03009} {\bibfield  {journal} {\bibinfo
  {journal} {Nano Letters}\ }\textbf {\bibinfo {volume} {15}},\ \bibinfo
  {pages} {7002} (\bibinfo {year} {2015})},\ \bibinfo {note} {pMID: 26422057},\
  \Eprint {http://arxiv.org/abs/https://doi.org/10.1021/acs.nanolett.5b03009}
  {https://doi.org/10.1021/acs.nanolett.5b03009} \BibitemShut {NoStop}%
\bibitem [{\citenamefont {Kidd}\ \emph {et~al.}(2016)\citenamefont {Kidd},
  \citenamefont {Zhang},\ and\ \citenamefont {Varga}}]{PhysRevB.93.125423}%
  \BibitemOpen
  \bibfield  {author} {\bibinfo {author} {\bibfnamefont {D.~W.}\ \bibnamefont
  {Kidd}}, \bibinfo {author} {\bibfnamefont {D.~K.}\ \bibnamefont {Zhang}}, \
  and\ \bibinfo {author} {\bibfnamefont {K.}~\bibnamefont {Varga}},\ }\href
  {\doibase 10.1103/PhysRevB.93.125423} {\bibfield  {journal} {\bibinfo
  {journal} {Phys. Rev. B}\ }\textbf {\bibinfo {volume} {93}},\ \bibinfo
  {pages} {125423} (\bibinfo {year} {2016})}\BibitemShut {NoStop}%
\bibitem [{\citenamefont {Drüppel}\ \emph {et~al.}(2017)\citenamefont
  {Drüppel}, \citenamefont {Deilmann}, \citenamefont {Krüger},\ and\
  \citenamefont {Rohlfing}}]{Drueppel2017}%
  \BibitemOpen
  \bibfield  {author} {\bibinfo {author} {\bibfnamefont {M.}~\bibnamefont
  {Drüppel}}, \bibinfo {author} {\bibfnamefont {T.}~\bibnamefont {Deilmann}},
  \bibinfo {author} {\bibfnamefont {P.}~\bibnamefont {Krüger}}, \ and\
  \bibinfo {author} {\bibfnamefont {M.}~\bibnamefont {Rohlfing}},\ }\href@noop
  {} {\bibfield  {journal} {\bibinfo  {journal} {Nature Communications}\
  }\textbf {\bibinfo {volume} {8}},\ \bibinfo {pages} {2117} (\bibinfo {year}
  {2017})}\BibitemShut {NoStop}%
\bibitem [{\citenamefont {Steinhoff}\ \emph
  {et~al.}(2018{\natexlab{b}})\citenamefont {Steinhoff}, \citenamefont
  {Florian}, \citenamefont {Singh}, \citenamefont {Tran}, \citenamefont
  {Kolarczik}, \citenamefont {Helmrich}, \citenamefont {Achtstein},
  \citenamefont {Woggon}, \citenamefont {Owschimikow}, \citenamefont {Jahnke},\
  and\ \citenamefont {Li}}]{steinhoff2018biexciton}%
  \BibitemOpen
  \bibfield  {author} {\bibinfo {author} {\bibfnamefont {A.}~\bibnamefont
  {Steinhoff}}, \bibinfo {author} {\bibfnamefont {M.}~\bibnamefont {Florian}},
  \bibinfo {author} {\bibfnamefont {A.}~\bibnamefont {Singh}}, \bibinfo
  {author} {\bibfnamefont {K.}~\bibnamefont {Tran}}, \bibinfo {author}
  {\bibfnamefont {M.}~\bibnamefont {Kolarczik}}, \bibinfo {author}
  {\bibfnamefont {S.}~\bibnamefont {Helmrich}}, \bibinfo {author}
  {\bibfnamefont {A.~W.}\ \bibnamefont {Achtstein}}, \bibinfo {author}
  {\bibfnamefont {U.}~\bibnamefont {Woggon}}, \bibinfo {author} {\bibfnamefont
  {N.}~\bibnamefont {Owschimikow}}, \bibinfo {author} {\bibfnamefont
  {F.}~\bibnamefont {Jahnke}}, \ and\ \bibinfo {author} {\bibfnamefont
  {X.}~\bibnamefont {Li}},\ }\href {\doibase
  https://doi.org/10.1038/s41567-018-0282-x} {\bibfield  {journal} {\bibinfo
  {journal} {Nature Physics}\ }\textbf {\bibinfo {volume} {14}},\ \bibinfo
  {pages} {1199} (\bibinfo {year} {2018}{\natexlab{b}})}\BibitemShut {NoStop}%
\bibitem [{\citenamefont {Katsch}\ \emph {et~al.}(2019)\citenamefont {Katsch},
  \citenamefont {Selig},\ and\ \citenamefont {Knorr}}]{katsch2019theory}%
  \BibitemOpen
  \bibfield  {author} {\bibinfo {author} {\bibfnamefont {F.}~\bibnamefont
  {Katsch}}, \bibinfo {author} {\bibfnamefont {M.}~\bibnamefont {Selig}}, \
  and\ \bibinfo {author} {\bibfnamefont {A.}~\bibnamefont {Knorr}},\
  }\href@noop {} {\bibfield  {journal} {\bibinfo  {journal} {2D Materials}\
  }\textbf {\bibinfo {volume} {7}},\ \bibinfo {pages} {015021} (\bibinfo {year}
  {2019})}\BibitemShut {NoStop}%
\bibitem [{\citenamefont {Kuhn}\ and\ \citenamefont
  {Richter}(2019)}]{Kuhn:2019}%
  \BibitemOpen
  \bibfield  {author} {\bibinfo {author} {\bibfnamefont {S.~C.}\ \bibnamefont
  {Kuhn}}\ and\ \bibinfo {author} {\bibfnamefont {M.}~\bibnamefont {Richter}},\
  }\href {\doibase 10.1103/PhysRevB.99.241301} {\bibfield  {journal} {\bibinfo
  {journal} {Phys. Rev. B}\ }\textbf {\bibinfo {volume} {99}},\ \bibinfo
  {pages} {241301} (\bibinfo {year} {2019})}\BibitemShut {NoStop}%
\bibitem [{\citenamefont {Vidal}(2003)}]{vidal2003efficient}%
  \BibitemOpen
  \bibfield  {author} {\bibinfo {author} {\bibfnamefont {G.}~\bibnamefont
  {Vidal}},\ }\href@noop {} {\bibfield  {journal} {\bibinfo  {journal} {Phys.
  Rev. Lett.}\ }\textbf {\bibinfo {volume} {91}},\ \bibinfo {pages} {147902}
  (\bibinfo {year} {2003})}\BibitemShut {NoStop}%
\bibitem [{\citenamefont {Or{\'u}s}(2014)}]{orus2014practical}%
  \BibitemOpen
  \bibfield  {author} {\bibinfo {author} {\bibfnamefont {R.}~\bibnamefont
  {Or{\'u}s}},\ }\href@noop {} {\bibfield  {journal} {\bibinfo  {journal}
  {Annals of Physics}\ }\textbf {\bibinfo {volume} {349}},\ \bibinfo {pages}
  {117} (\bibinfo {year} {2014})}\BibitemShut {NoStop}%
\bibitem [{\citenamefont {Schollw{\"o}ck}(2011)}]{schollwock2011density}%
  \BibitemOpen
  \bibfield  {author} {\bibinfo {author} {\bibfnamefont {U.}~\bibnamefont
  {Schollw{\"o}ck}},\ }\href@noop {} {\bibfield  {journal} {\bibinfo  {journal}
  {Annals of Physics}\ }\textbf {\bibinfo {volume} {326}},\ \bibinfo {pages}
  {96} (\bibinfo {year} {2011})}\BibitemShut {NoStop}%
\bibitem [{\citenamefont {Cirac}\ \emph {et~al.}(2017)\citenamefont {Cirac},
  \citenamefont {Pérez-García}, \citenamefont {Schuch},\ and\ \citenamefont
  {Verstraete}}]{CIRAC2017100}%
  \BibitemOpen
  \bibfield  {author} {\bibinfo {author} {\bibfnamefont {J.}~\bibnamefont
  {Cirac}}, \bibinfo {author} {\bibfnamefont {D.}~\bibnamefont
  {Pérez-García}}, \bibinfo {author} {\bibfnamefont {N.}~\bibnamefont
  {Schuch}}, \ and\ \bibinfo {author} {\bibfnamefont {F.}~\bibnamefont
  {Verstraete}},\ }\href {\doibase https://doi.org/10.1016/j.aop.2016.12.030}
  {\bibfield  {journal} {\bibinfo  {journal} {Annals of Physics}\ }\textbf
  {\bibinfo {volume} {378}},\ \bibinfo {pages} {100 } (\bibinfo {year}
  {2017})}\BibitemShut {NoStop}%
\bibitem [{\citenamefont {Verstraete}\ and\ \citenamefont
  {Cirac}(2006)}]{verstraete2006matrix}%
  \BibitemOpen
  \bibfield  {author} {\bibinfo {author} {\bibfnamefont {F.}~\bibnamefont
  {Verstraete}}\ and\ \bibinfo {author} {\bibfnamefont {J.~I.}\ \bibnamefont
  {Cirac}},\ }\href@noop {} {\bibfield  {journal} {\bibinfo  {journal}
  {Physical Review B}\ }\textbf {\bibinfo {volume} {73}},\ \bibinfo {pages}
  {094423} (\bibinfo {year} {2006})}\BibitemShut {NoStop}%
\bibitem [{\citenamefont {Vidal}(2007)}]{vidal2007classical}%
  \BibitemOpen
  \bibfield  {author} {\bibinfo {author} {\bibfnamefont {G.}~\bibnamefont
  {Vidal}},\ }\href@noop {} {\bibfield  {journal} {\bibinfo  {journal}
  {Physical review letters}\ }\textbf {\bibinfo {volume} {98}},\ \bibinfo
  {pages} {070201} (\bibinfo {year} {2007})}\BibitemShut {NoStop}%
\bibitem [{\citenamefont {Clark}\ \emph {et~al.}(2010)\citenamefont {Clark},
  \citenamefont {Prior}, \citenamefont {Hartmann}, \citenamefont {Jaksch},\
  and\ \citenamefont {Plenio}}]{plenioheisenberg}%
  \BibitemOpen
  \bibfield  {author} {\bibinfo {author} {\bibfnamefont {S.~R.}\ \bibnamefont
  {Clark}}, \bibinfo {author} {\bibfnamefont {J.}~\bibnamefont {Prior}},
  \bibinfo {author} {\bibfnamefont {M.~J.}\ \bibnamefont {Hartmann}}, \bibinfo
  {author} {\bibfnamefont {D.}~\bibnamefont {Jaksch}}, \ and\ \bibinfo {author}
  {\bibfnamefont {M.~B.}\ \bibnamefont {Plenio}},\ }\href
  {http://stacks.iop.org/1367-2630/12/i=2/a=025005} {\bibfield  {journal}
  {\bibinfo  {journal} {New Journal of Physics}\ }\textbf {\bibinfo {volume}
  {12}},\ \bibinfo {pages} {025005} (\bibinfo {year} {2010})}\BibitemShut
  {NoStop}%
\bibitem [{\citenamefont {White}(1992)}]{white1992density}%
  \BibitemOpen
  \bibfield  {author} {\bibinfo {author} {\bibfnamefont {S.~R.}\ \bibnamefont
  {White}},\ }\href@noop {} {\bibfield  {journal} {\bibinfo  {journal}
  {Physical review letters}\ }\textbf {\bibinfo {volume} {69}},\ \bibinfo
  {pages} {2863} (\bibinfo {year} {1992})}\BibitemShut {NoStop}%
\bibitem [{\citenamefont {Oseledets}(2009)}]{oseledets2009approximation}%
  \BibitemOpen
  \bibfield  {author} {\bibinfo {author} {\bibfnamefont {I.}~\bibnamefont
  {Oseledets}},\ }in\ \href@noop {} {\emph {\bibinfo {booktitle} {Doklady
  Mathematics}}},\ Vol.~\bibinfo {volume} {80}\ (\bibinfo {organization}
  {Springer},\ \bibinfo {year} {2009})\ pp.\ \bibinfo {pages}
  {653--654}\BibitemShut {NoStop}%
\bibitem [{\citenamefont {Oseledets}(2010)}]{oseledets2010approximation}%
  \BibitemOpen
  \bibfield  {author} {\bibinfo {author} {\bibfnamefont {I.~V.}\ \bibnamefont
  {Oseledets}},\ }\href@noop {} {\bibfield  {journal} {\bibinfo  {journal}
  {SIAM Journal on Matrix Analysis and Applications}\ }\textbf {\bibinfo
  {volume} {31}},\ \bibinfo {pages} {2130} (\bibinfo {year}
  {2010})}\BibitemShut {NoStop}%
\bibitem [{\citenamefont {Khoromskij}(2011)}]{khoromskij2011dlog}%
  \BibitemOpen
  \bibfield  {author} {\bibinfo {author} {\bibfnamefont {B.~N.}\ \bibnamefont
  {Khoromskij}},\ }\href@noop {} {\bibfield  {journal} {\bibinfo  {journal}
  {Constructive Approximation}\ }\textbf {\bibinfo {volume} {34}},\ \bibinfo
  {pages} {257} (\bibinfo {year} {2011})}\BibitemShut {NoStop}%
\bibitem [{\citenamefont {Kazeev}\ and\ \citenamefont
  {Khoromskij}(2012)}]{kazeev2012low}%
  \BibitemOpen
  \bibfield  {author} {\bibinfo {author} {\bibfnamefont {V.~A.}\ \bibnamefont
  {Kazeev}}\ and\ \bibinfo {author} {\bibfnamefont {B.~N.}\ \bibnamefont
  {Khoromskij}},\ }\href@noop {} {\bibfield  {journal} {\bibinfo  {journal}
  {SIAM J. Matrix Anal. Appl.}\ }\textbf {\bibinfo {volume} {33}},\ \bibinfo
  {pages} {742} (\bibinfo {year} {2012})}\BibitemShut {NoStop}%
\bibitem [{\citenamefont {Khoromskaia}\ and\ \citenamefont
  {Khoromskij}(2015)}]{khoromskaia2015tensor}%
  \BibitemOpen
  \bibfield  {author} {\bibinfo {author} {\bibfnamefont {V.}~\bibnamefont
  {Khoromskaia}}\ and\ \bibinfo {author} {\bibfnamefont {B.~N.}\ \bibnamefont
  {Khoromskij}},\ }\href@noop {} {\bibfield  {journal} {\bibinfo  {journal}
  {Physical Chemistry Chemical Physics}\ }\textbf {\bibinfo {volume} {17}},\
  \bibinfo {pages} {31491} (\bibinfo {year} {2015})}\BibitemShut {NoStop}%
\bibitem [{\citenamefont {Benner}\ \emph {et~al.}(2017)\citenamefont {Benner},
  \citenamefont {Dolgov}, \citenamefont {Khoromskaia},\ and\ \citenamefont
  {Khoromskij}}]{benner2017fast}%
  \BibitemOpen
  \bibfield  {author} {\bibinfo {author} {\bibfnamefont {P.}~\bibnamefont
  {Benner}}, \bibinfo {author} {\bibfnamefont {S.}~\bibnamefont {Dolgov}},
  \bibinfo {author} {\bibfnamefont {V.}~\bibnamefont {Khoromskaia}}, \ and\
  \bibinfo {author} {\bibfnamefont {B.~N.}\ \bibnamefont {Khoromskij}},\
  }\href@noop {} {\bibfield  {journal} {\bibinfo  {journal} {Journal of
  Computational Physics}\ }\textbf {\bibinfo {volume} {334}},\ \bibinfo {pages}
  {221} (\bibinfo {year} {2017})}\BibitemShut {NoStop}%
\bibitem [{\citenamefont {Ridolfi}\ \emph {et~al.}(2015)\citenamefont
  {Ridolfi}, \citenamefont {Le}, \citenamefont {Rahman}, \citenamefont
  {Mucciolo},\ and\ \citenamefont {Lewenkopf}}]{ridolfi2015tight}%
  \BibitemOpen
  \bibfield  {author} {\bibinfo {author} {\bibfnamefont {E.}~\bibnamefont
  {Ridolfi}}, \bibinfo {author} {\bibfnamefont {D.}~\bibnamefont {Le}},
  \bibinfo {author} {\bibfnamefont {T.}~\bibnamefont {Rahman}}, \bibinfo
  {author} {\bibfnamefont {E.}~\bibnamefont {Mucciolo}}, \ and\ \bibinfo
  {author} {\bibfnamefont {C.}~\bibnamefont {Lewenkopf}},\ }\href@noop {}
  {\bibfield  {journal} {\bibinfo  {journal} {Journal of Physics: Condensed
  Matter}\ }\textbf {\bibinfo {volume} {27}},\ \bibinfo {pages} {365501}
  (\bibinfo {year} {2015})}\BibitemShut {NoStop}%
\bibitem [{\citenamefont {Berkelbach}\ \emph {et~al.}(2013)\citenamefont
  {Berkelbach}, \citenamefont {Hybertsen},\ and\ \citenamefont
  {Reichman}}]{PhysRevB.88.045318}%
  \BibitemOpen
  \bibfield  {author} {\bibinfo {author} {\bibfnamefont {T.~C.}\ \bibnamefont
  {Berkelbach}}, \bibinfo {author} {\bibfnamefont {M.~S.}\ \bibnamefont
  {Hybertsen}}, \ and\ \bibinfo {author} {\bibfnamefont {D.~R.}\ \bibnamefont
  {Reichman}},\ }\href {\doibase 10.1103/PhysRevB.88.045318} {\bibfield
  {journal} {\bibinfo  {journal} {Phys. Rev. B}\ }\textbf {\bibinfo {volume}
  {88}},\ \bibinfo {pages} {045318} (\bibinfo {year} {2013})}\BibitemShut
  {NoStop}%
\bibitem [{\citenamefont {Takagahara}(1993)}]{PhysRevB.47.4569}%
  \BibitemOpen
  \bibfield  {author} {\bibinfo {author} {\bibfnamefont {T.}~\bibnamefont
  {Takagahara}},\ }\href {\doibase 10.1103/PhysRevB.47.4569} {\bibfield
  {journal} {\bibinfo  {journal} {Phys. Rev. B}\ }\textbf {\bibinfo {volume}
  {47}},\ \bibinfo {pages} {4569} (\bibinfo {year} {1993})}\BibitemShut
  {NoStop}%
\bibitem [{\citenamefont {Marzari}\ \emph {et~al.}(2012)\citenamefont
  {Marzari}, \citenamefont {Mostofi}, \citenamefont {Yates}, \citenamefont
  {Souza},\ and\ \citenamefont {Vanderbilt}}]{RevModPhys.84.1419}%
  \BibitemOpen
  \bibfield  {author} {\bibinfo {author} {\bibfnamefont {N.}~\bibnamefont
  {Marzari}}, \bibinfo {author} {\bibfnamefont {A.~A.}\ \bibnamefont
  {Mostofi}}, \bibinfo {author} {\bibfnamefont {J.~R.}\ \bibnamefont {Yates}},
  \bibinfo {author} {\bibfnamefont {I.}~\bibnamefont {Souza}}, \ and\ \bibinfo
  {author} {\bibfnamefont {D.}~\bibnamefont {Vanderbilt}},\ }\href {\doibase
  10.1103/RevModPhys.84.1419} {\bibfield  {journal} {\bibinfo  {journal} {Rev.
  Mod. Phys.}\ }\textbf {\bibinfo {volume} {84}},\ \bibinfo {pages} {1419}
  (\bibinfo {year} {2012})}\BibitemShut {NoStop}%
\bibitem [{\citenamefont {Mostofi}\ \emph {et~al.}(2008)\citenamefont
  {Mostofi}, \citenamefont {Yates}, \citenamefont {Lee}, \citenamefont {Souza},
  \citenamefont {Vanderbilt},\ and\ \citenamefont {Marzari}}]{MOSTOFI2008685}%
  \BibitemOpen
  \bibfield  {author} {\bibinfo {author} {\bibfnamefont {A.~A.}\ \bibnamefont
  {Mostofi}}, \bibinfo {author} {\bibfnamefont {J.~R.}\ \bibnamefont {Yates}},
  \bibinfo {author} {\bibfnamefont {Y.-S.}\ \bibnamefont {Lee}}, \bibinfo
  {author} {\bibfnamefont {I.}~\bibnamefont {Souza}}, \bibinfo {author}
  {\bibfnamefont {D.}~\bibnamefont {Vanderbilt}}, \ and\ \bibinfo {author}
  {\bibfnamefont {N.}~\bibnamefont {Marzari}},\ }\href {\doibase
  https://doi.org/10.1016/j.cpc.2007.11.016} {\bibfield  {journal} {\bibinfo
  {journal} {Computer Physics Communications}\ }\textbf {\bibinfo {volume}
  {178}},\ \bibinfo {pages} {685 } (\bibinfo {year} {2008})}\BibitemShut
  {NoStop}%
\bibitem [{\citenamefont {Hao}\ \emph {et~al.}(2017)\citenamefont {Hao},
  \citenamefont {Specht}, \citenamefont {Nagler}, \citenamefont {Xu},
  \citenamefont {Tran}, \citenamefont {Singh}, \citenamefont {Dass},
  \citenamefont {Sch{\"u}ller}, \citenamefont {Korn}, \citenamefont {Richter}
  \emph {et~al.}}]{hao2017neutral}%
  \BibitemOpen
  \bibfield  {author} {\bibinfo {author} {\bibfnamefont {K.}~\bibnamefont
  {Hao}}, \bibinfo {author} {\bibfnamefont {J.~F.}\ \bibnamefont {Specht}},
  \bibinfo {author} {\bibfnamefont {P.}~\bibnamefont {Nagler}}, \bibinfo
  {author} {\bibfnamefont {L.}~\bibnamefont {Xu}}, \bibinfo {author}
  {\bibfnamefont {K.}~\bibnamefont {Tran}}, \bibinfo {author} {\bibfnamefont
  {A.}~\bibnamefont {Singh}}, \bibinfo {author} {\bibfnamefont {C.~K.}\
  \bibnamefont {Dass}}, \bibinfo {author} {\bibfnamefont {C.}~\bibnamefont
  {Sch{\"u}ller}}, \bibinfo {author} {\bibfnamefont {T.}~\bibnamefont {Korn}},
  \bibinfo {author} {\bibfnamefont {M.}~\bibnamefont {Richter}},  \emph
  {et~al.},\ }\href@noop {} {\bibfield  {journal} {\bibinfo  {journal} {Nature
  communications}\ }\textbf {\bibinfo {volume} {8}},\ \bibinfo {pages} {15552}
  (\bibinfo {year} {2017})}\BibitemShut {NoStop}%
\bibitem [{\citenamefont {Monkhorst}\ and\ \citenamefont
  {Pack}(1976)}]{PhysRevB.13.5188}%
  \BibitemOpen
  \bibfield  {author} {\bibinfo {author} {\bibfnamefont {H.~J.}\ \bibnamefont
  {Monkhorst}}\ and\ \bibinfo {author} {\bibfnamefont {J.~D.}\ \bibnamefont
  {Pack}},\ }\href {\doibase 10.1103/PhysRevB.13.5188} {\bibfield  {journal}
  {\bibinfo  {journal} {Phys. Rev. B}\ }\textbf {\bibinfo {volume} {13}},\
  \bibinfo {pages} {5188} (\bibinfo {year} {1976})}\BibitemShut {NoStop}%
\bibitem [{\citenamefont {Eisert}\ \emph {et~al.}(2010)\citenamefont {Eisert},
  \citenamefont {Cramer},\ and\ \citenamefont {Plenio}}]{RevModPhys.82.277}%
  \BibitemOpen
  \bibfield  {author} {\bibinfo {author} {\bibfnamefont {J.}~\bibnamefont
  {Eisert}}, \bibinfo {author} {\bibfnamefont {M.}~\bibnamefont {Cramer}}, \
  and\ \bibinfo {author} {\bibfnamefont {M.~B.}\ \bibnamefont {Plenio}},\
  }\href {\doibase 10.1103/RevModPhys.82.277} {\bibfield  {journal} {\bibinfo
  {journal} {Rev. Mod. Phys.}\ }\textbf {\bibinfo {volume} {82}},\ \bibinfo
  {pages} {277} (\bibinfo {year} {2010})}\BibitemShut {NoStop}%
\bibitem [{\citenamefont {Tietze}\ \emph {et~al.}(2015)\citenamefont {Tietze},
  \citenamefont {Schenk},\ and\ \citenamefont {Gamm}}]{tietze2015electronic}%
  \BibitemOpen
  \bibfield  {author} {\bibinfo {author} {\bibfnamefont {U.}~\bibnamefont
  {Tietze}}, \bibinfo {author} {\bibfnamefont {C.}~\bibnamefont {Schenk}}, \
  and\ \bibinfo {author} {\bibfnamefont {E.}~\bibnamefont {Gamm}},\ }\href@noop
  {} {\emph {\bibinfo {title} {Electronic circuits: handbook for design and
  application}}}\ (\bibinfo  {publisher} {Springer},\ \bibinfo {year}
  {2015})\BibitemShut {NoStop}%
\bibitem [{\citenamefont {Stoudenmire}\ and\ \citenamefont
  {White}()}]{itensor}%
  \BibitemOpen
  \bibfield  {author} {\bibinfo {author} {\bibfnamefont {E.~M.}\ \bibnamefont
  {Stoudenmire}}\ and\ \bibinfo {author} {\bibfnamefont {S.~R.}\ \bibnamefont
  {White}},\ }\href {http://itensor.org/} {\enquote {\bibinfo {title} {Itensor
  c++ library,http://itensor.org/},}\ }\BibitemShut {NoStop}%
\bibitem [{\citenamefont {Lim}\ and\ \citenamefont
  {Sheng}(2016)}]{PhysRevB.94.045111}%
  \BibitemOpen
  \bibfield  {author} {\bibinfo {author} {\bibfnamefont {S.~P.}\ \bibnamefont
  {Lim}}\ and\ \bibinfo {author} {\bibfnamefont {D.~N.}\ \bibnamefont
  {Sheng}},\ }\href {\doibase 10.1103/PhysRevB.94.045111} {\bibfield  {journal}
  {\bibinfo  {journal} {Phys. Rev. B}\ }\textbf {\bibinfo {volume} {94}},\
  \bibinfo {pages} {045111} (\bibinfo {year} {2016})}\BibitemShut {NoStop}%
\bibitem [{\citenamefont {Yu}\ \emph {et~al.}(2017)\citenamefont {Yu},
  \citenamefont {Pekker},\ and\ \citenamefont
  {Clark}}]{PhysRevLett.118.017201}%
  \BibitemOpen
  \bibfield  {author} {\bibinfo {author} {\bibfnamefont {X.}~\bibnamefont
  {Yu}}, \bibinfo {author} {\bibfnamefont {D.}~\bibnamefont {Pekker}}, \ and\
  \bibinfo {author} {\bibfnamefont {B.~K.}\ \bibnamefont {Clark}},\ }\href
  {\doibase 10.1103/PhysRevLett.118.017201} {\bibfield  {journal} {\bibinfo
  {journal} {Phys. Rev. Lett.}\ }\textbf {\bibinfo {volume} {118}},\ \bibinfo
  {pages} {017201} (\bibinfo {year} {2017})}\BibitemShut {NoStop}%
\bibitem [{\citenamefont {Khemani}\ \emph {et~al.}(2016)\citenamefont
  {Khemani}, \citenamefont {Pollmann},\ and\ \citenamefont
  {Sondhi}}]{PhysRevLett.116.247204}%
  \BibitemOpen
  \bibfield  {author} {\bibinfo {author} {\bibfnamefont {V.}~\bibnamefont
  {Khemani}}, \bibinfo {author} {\bibfnamefont {F.}~\bibnamefont {Pollmann}}, \
  and\ \bibinfo {author} {\bibfnamefont {S.~L.}\ \bibnamefont {Sondhi}},\
  }\href {\doibase 10.1103/PhysRevLett.116.247204} {\bibfield  {journal}
  {\bibinfo  {journal} {Phys. Rev. Lett.}\ }\textbf {\bibinfo {volume} {116}},\
  \bibinfo {pages} {247204} (\bibinfo {year} {2016})}\BibitemShut {NoStop}%
\bibitem [{\citenamefont {Richter}(2017)}]{richter2017nanoplatelets}%
  \BibitemOpen
  \bibfield  {author} {\bibinfo {author} {\bibfnamefont {M.}~\bibnamefont
  {Richter}},\ }\href@noop {} {\bibfield  {journal} {\bibinfo  {journal}
  {Physical Review Materials}\ }\textbf {\bibinfo {volume} {1}},\ \bibinfo
  {pages} {016001} (\bibinfo {year} {2017})}\BibitemShut {NoStop}%
\bibitem [{\citenamefont {Specht}\ \emph {et~al.}(2019)\citenamefont {Specht},
  \citenamefont {Scott}, \citenamefont {Corona~Castro}, \citenamefont
  {Christodoulou}, \citenamefont {Bertrand}, \citenamefont {Prudnikau},
  \citenamefont {Antanovich}, \citenamefont {Siebbeles}, \citenamefont
  {Owschimikow}, \citenamefont {Moreels}, \citenamefont {Artemyev},
  \citenamefont {Woggon}, \citenamefont {Achtstein},\ and\ \citenamefont
  {Richter}}]{C9NR03161H}%
  \BibitemOpen
  \bibfield  {author} {\bibinfo {author} {\bibfnamefont {J.~F.}\ \bibnamefont
  {Specht}}, \bibinfo {author} {\bibfnamefont {R.}~\bibnamefont {Scott}},
  \bibinfo {author} {\bibfnamefont {M.}~\bibnamefont {Corona~Castro}}, \bibinfo
  {author} {\bibfnamefont {S.}~\bibnamefont {Christodoulou}}, \bibinfo {author}
  {\bibfnamefont {G.~H.~V.}\ \bibnamefont {Bertrand}}, \bibinfo {author}
  {\bibfnamefont {A.~V.}\ \bibnamefont {Prudnikau}}, \bibinfo {author}
  {\bibfnamefont {A.}~\bibnamefont {Antanovich}}, \bibinfo {author}
  {\bibfnamefont {L.~D.~A.}\ \bibnamefont {Siebbeles}}, \bibinfo {author}
  {\bibfnamefont {N.}~\bibnamefont {Owschimikow}}, \bibinfo {author}
  {\bibfnamefont {I.}~\bibnamefont {Moreels}}, \bibinfo {author} {\bibfnamefont
  {M.}~\bibnamefont {Artemyev}}, \bibinfo {author} {\bibfnamefont
  {U.}~\bibnamefont {Woggon}}, \bibinfo {author} {\bibfnamefont {A.~W.}\
  \bibnamefont {Achtstein}}, \ and\ \bibinfo {author} {\bibfnamefont
  {M.}~\bibnamefont {Richter}},\ }\href {\doibase 10.1039/C9NR03161H}
  {\bibfield  {journal} {\bibinfo  {journal} {Nanoscale}\ }\textbf {\bibinfo
  {volume} {11}},\ \bibinfo {pages} {12230} (\bibinfo {year}
  {2019})}\BibitemShut {NoStop}%
\bibitem [{\citenamefont {Qiu}\ \emph {et~al.}(2013{\natexlab{b}})\citenamefont
  {Qiu}, \citenamefont {da~Jornada},\ and\ \citenamefont
  {Louie}}]{PhysRevLett.111.216805}%
  \BibitemOpen
  \bibfield  {author} {\bibinfo {author} {\bibfnamefont {D.~Y.}\ \bibnamefont
  {Qiu}}, \bibinfo {author} {\bibfnamefont {F.~H.}\ \bibnamefont {da~Jornada}},
  \ and\ \bibinfo {author} {\bibfnamefont {S.~G.}\ \bibnamefont {Louie}},\
  }\href {\doibase 10.1103/PhysRevLett.111.216805} {\bibfield  {journal}
  {\bibinfo  {journal} {Phys. Rev. Lett.}\ }\textbf {\bibinfo {volume} {111}},\
  \bibinfo {pages} {216805} (\bibinfo {year} {2013}{\natexlab{b}})}\BibitemShut
  {NoStop}%
\bibitem [{\citenamefont {Bergh{\"a}user}\ \emph {et~al.}(2018)\citenamefont
  {Bergh{\"a}user}, \citenamefont {Bernal-Villamil}, \citenamefont {Schmidt},
  \citenamefont {Schneider}, \citenamefont {Niehues}, \citenamefont {Erhart},
  \citenamefont {de~Vasconcellos}, \citenamefont {Bratschitsch}, \citenamefont
  {Knorr},\ and\ \citenamefont {Malic}}]{berghauser2018inverted}%
  \BibitemOpen
  \bibfield  {author} {\bibinfo {author} {\bibfnamefont {G.}~\bibnamefont
  {Bergh{\"a}user}}, \bibinfo {author} {\bibfnamefont {I.}~\bibnamefont
  {Bernal-Villamil}}, \bibinfo {author} {\bibfnamefont {R.}~\bibnamefont
  {Schmidt}}, \bibinfo {author} {\bibfnamefont {R.}~\bibnamefont {Schneider}},
  \bibinfo {author} {\bibfnamefont {I.}~\bibnamefont {Niehues}}, \bibinfo
  {author} {\bibfnamefont {P.}~\bibnamefont {Erhart}}, \bibinfo {author}
  {\bibfnamefont {S.~M.}\ \bibnamefont {de~Vasconcellos}}, \bibinfo {author}
  {\bibfnamefont {R.}~\bibnamefont {Bratschitsch}}, \bibinfo {author}
  {\bibfnamefont {A.}~\bibnamefont {Knorr}}, \ and\ \bibinfo {author}
  {\bibfnamefont {E.}~\bibnamefont {Malic}},\ }\href@noop {} {\bibfield
  {journal} {\bibinfo  {journal} {Nature communications}\ }\textbf {\bibinfo
  {volume} {9}},\ \bibinfo {pages} {971} (\bibinfo {year} {2018})}\BibitemShut
  {NoStop}%
\bibitem [{\citenamefont {Selig}\ \emph {et~al.}(2018)\citenamefont {Selig},
  \citenamefont {Bergh{\"a}user}, \citenamefont {Richter}, \citenamefont
  {Bratschitsch}, \citenamefont {Knorr},\ and\ \citenamefont
  {Malic}}]{maltedarkbright}%
  \BibitemOpen
  \bibfield  {author} {\bibinfo {author} {\bibfnamefont {M.}~\bibnamefont
  {Selig}}, \bibinfo {author} {\bibfnamefont {G.}~\bibnamefont
  {Bergh{\"a}user}}, \bibinfo {author} {\bibfnamefont {M.}~\bibnamefont
  {Richter}}, \bibinfo {author} {\bibfnamefont {R.}~\bibnamefont
  {Bratschitsch}}, \bibinfo {author} {\bibfnamefont {A.}~\bibnamefont {Knorr}},
  \ and\ \bibinfo {author} {\bibfnamefont {E.}~\bibnamefont {Malic}},\ }\href
  {http://stacks.iop.org/2053-1583/5/i=3/a=035017} {\bibfield  {journal}
  {\bibinfo  {journal} {2D Materials}\ }\textbf {\bibinfo {volume} {5}},\
  \bibinfo {pages} {035017} (\bibinfo {year} {2018})}\BibitemShut {NoStop}%
\bibitem [{\citenamefont {Peeters}\ \emph {et~al.}(2001)\citenamefont
  {Peeters}, \citenamefont {Riva},\ and\ \citenamefont
  {Varga}}]{PEETERS2001139}%
  \BibitemOpen
  \bibfield  {author} {\bibinfo {author} {\bibfnamefont {F.}~\bibnamefont
  {Peeters}}, \bibinfo {author} {\bibfnamefont {C.}~\bibnamefont {Riva}}, \
  and\ \bibinfo {author} {\bibfnamefont {K.}~\bibnamefont {Varga}},\ }\href
  {\doibase https://doi.org/10.1016/S0921-4526(01)00577-4} {\bibfield
  {journal} {\bibinfo  {journal} {Physica B: Condensed Matter}\ }\textbf
  {\bibinfo {volume} {300}},\ \bibinfo {pages} {139 } (\bibinfo {year}
  {2001})},\ \bibinfo {note} {jubilee issue Volume 300}\BibitemShut {NoStop}%
\bibitem [{\citenamefont {Palacios}\ \emph {et~al.}(1996)\citenamefont
  {Palacios}, \citenamefont {Yoshioka},\ and\ \citenamefont
  {MacDonald}}]{PhysRevB.54.R2296}%
  \BibitemOpen
  \bibfield  {author} {\bibinfo {author} {\bibfnamefont {J.~J.}\ \bibnamefont
  {Palacios}}, \bibinfo {author} {\bibfnamefont {D.}~\bibnamefont {Yoshioka}},
  \ and\ \bibinfo {author} {\bibfnamefont {A.~H.}\ \bibnamefont {MacDonald}},\
  }\href {\doibase 10.1103/PhysRevB.54.R2296} {\bibfield  {journal} {\bibinfo
  {journal} {Phys. Rev. B}\ }\textbf {\bibinfo {volume} {54}},\ \bibinfo
  {pages} {R2296} (\bibinfo {year} {1996})}\BibitemShut {NoStop}%
\bibitem [{\citenamefont {Szyniszewski}\ \emph {et~al.}(2017)\citenamefont
  {Szyniszewski}, \citenamefont {Mostaani}, \citenamefont {Drummond},\ and\
  \citenamefont {Fal'ko}}]{PhysRevB.95.081301}%
  \BibitemOpen
  \bibfield  {author} {\bibinfo {author} {\bibfnamefont {M.}~\bibnamefont
  {Szyniszewski}}, \bibinfo {author} {\bibfnamefont {E.}~\bibnamefont
  {Mostaani}}, \bibinfo {author} {\bibfnamefont {N.~D.}\ \bibnamefont
  {Drummond}}, \ and\ \bibinfo {author} {\bibfnamefont {V.~I.}\ \bibnamefont
  {Fal'ko}},\ }\href {\doibase 10.1103/PhysRevB.95.081301} {\bibfield
  {journal} {\bibinfo  {journal} {Phys. Rev. B}\ }\textbf {\bibinfo {volume}
  {95}},\ \bibinfo {pages} {081301} (\bibinfo {year} {2017})}\BibitemShut
  {NoStop}%
\bibitem [{\citenamefont {Kyl\"anp\"a\"a}\ and\ \citenamefont
  {Komsa}(2015)}]{PhysRevB.92.205418}%
  \BibitemOpen
  \bibfield  {author} {\bibinfo {author} {\bibfnamefont {I.}~\bibnamefont
  {Kyl\"anp\"a\"a}}\ and\ \bibinfo {author} {\bibfnamefont {H.-P.}\
  \bibnamefont {Komsa}},\ }\href {\doibase 10.1103/PhysRevB.92.205418}
  {\bibfield  {journal} {\bibinfo  {journal} {Phys. Rev. B}\ }\textbf {\bibinfo
  {volume} {92}},\ \bibinfo {pages} {205418} (\bibinfo {year}
  {2015})}\BibitemShut {NoStop}%
\end{thebibliography}
%


\end{document}